\documentclass[10pt,aps,prb,showpacs,floatfix,raggedbottom,floats,superscriptaddress,twocolumn]{revtex4-1}
\pdfoutput=1
\usepackage{graphicx,latexsym}
\usepackage{dcolumn}
\usepackage{amssymb,amsmath,bm}
\usepackage{color}
\newcommand{\beq}{\begin{equation}}
\newcommand{\eeq}{\end{equation}}

\begin{document}

\title{Magneto-Optical Quantum Switching in a System of Spinor Excitons}

\author{Wen-Hsuan Kuan}
\email{wenhsuan.kuan@gmail.com}
\affiliation{Department of Applied Physics and Chemistry, University of Taipei, Taipei 10048, Taiwan.}
\author{Vidar Gudmundsson}
\email{vidar@hi.is}
\affiliation{Science Institute, University of Iceland, Dunhaga 3, IS-107 Reykjavik, Iceland}

\date{\today}

\begin{abstract}
       In this work we investigate magneto-optical {properties} of two-dimensional semiconductor quantum-ring excitons with Rashba and Dresselhaus spin-orbit interactions threaded by a magnetic flux perpendicular to the plane of the ring. By calculating the excitonic Aharonov-Bohm spectrum, we study the Coulomb and spin-orbit effects on the Aharonov-Bohm features. From the light-matter interactions of the excitons, we find that for scalar excitons, there are open channels for spontaneous recombination resulting in a bright photoluminescence spectrum, whereas the forbidden recombination of dipolar excitons results in a dark photoluminescence spectrum. We investigate the generation of persistent charge and spin currents. The exploration of spin orientations manifests that by adjusting the strength of the spin-orbit interactions, the exciton can be constructed as a squeezed complex with specific spin polarization. Moreover, a coherently moving dipolar exciton acquires a nontrivial dual Aharonov-Casher phase, creating the possibility to generate persistent dipole currents and spin dipole currents. Our study reveals that a manipulation of the spin-orbit interactions provides a potential application for quantum-ring spinor excitons to be utilized in nano-scaled magneto-optical switches.
\end{abstract}

\maketitle

\section{Introduction}

While the relativistic nature of a spin-orbit interaction (SOI) in atomic systems
is given to be in the order of $10^{-5}$\,eV,
a large SOI field can be obtained either in the presence of a large electric field or in materials where the mass gap is reduced.
Therefore, in recent decades, studies of spin-dependent effects in semiconductor nanostructures\cite{ManuelValin-Rodriguez2002,Meijer2002,Tsitsishvili2004,Kuan2004} and nano-devices\cite{Datta1990,Wang2002,Koga2002} have been attracted to both the experimental and the theoretical aspects and have opened up the field of
spintronics.\cite{Wolf2001,Ziutifmmodecuteclseci2004} In narrow-gap heterostructures
such as InAlAs/InGaAs, a spontaneous spin-splitting can be realized by structural inversion asymmetry (SIA).\cite{Rashba1960} The kind of asymmetry, corresponding to an inhomogeneous built-in electric field due to band-bending and discontinuity which gives rise to the 2DEG confining potential at the interface, leads to the splitting termed the Rashba effect.\cite{Bychkov1984} On the other hand, in wide-gap zinc-blend structures, bulk inversion asymmetry (BIA) may induce a coupling of electronic states which is cubic in the electronic momentum ${\bf k}$. The spin splitting of electron and hole states at nonzero ${\bf k}$, even at zero magnetic field, is known as the Dresselhaus effect.\cite{Dresselhaus1955}
Both spin splittings can be tuned continuously by means of external gates.

Achievements in the state-of-the-art material engineering and nanofabrication
techniques have successfully led to the realization of advanced semiconductor devices and made
possible the investigation of quantum phenomena in these systems.\cite{Fuhrer2001, Mano2005} In ring geometries, systems that undergo a slow, cyclic evolution were predicted
to induce a Berry phase for the electronic states.\cite{Berry1984}
The existence of the geometric quantum phase reveals the
significance of electromagnetic potentials in the quantum theory. The periodic interference effect identified by Aharonov and Bohm (AB)\cite{Aharonov1959} was measured via spectroscopy of nanoscopic semiconductor rings\cite{Lorke2000} and AB oscillations of the spin components were manifested in the resistance in a GaAs 2D hole system with a strong SOI.\cite{Yau2002} Many electronic properties, like optical absorption and transmission,\cite{Murdin1999} cyclotron-resonance,\cite{Krahne2001} and Rabi oscillations,\cite{Kamada2001,Stievater2001,Htoon2002,Borri2002,Stufler2005} have been examined in combination with the optical AB effect, with or without SOIs. However, the AB effect for another simple many-body complex, the exciton, has not caught the attention of researchers,
nor has its existence even been debated.

    Viewed as a neutral particle, the exciton was not expected to demonstrate
an AB effect. But it was pointed out that the optical AB effect could be manifested for polarized excitons in ideal nano rings,\cite{Govorov2002} and for Wigner molecules in Type-II quantum dots.\cite{Okuyama2011} Whenever there is a net magnetic flux traversing the particles, the electron-hole pair would be dragged along together, and their coherence would be revealed in an oscillatory energy spectrum. A robust AB oscillation
observed in columnar ZnTe/ZnSe quantum dots has provided the evidence to support
the formation of coherently rotating states of neutral excitons.\cite{Sellers2008,Roy2012}

   Based on previous experience, we proceed for more detailed and realistic
simulations by adopting a double-ring confining potential with a finite width. The geometry provides the degree of freedom to adjust the separation and the barrier between an electron and a hole, hence establishing an opportunity to determine the density correlations displaying the coherence properties of the system.
Meanwhile, the inclusion of the SOIs is aimed at testifying the AB stiffness under external perturbations. The model describes electrons in the $\Gamma$-6 band under the influence of the Rashba effect and heavy holes in the $\Gamma$-8 band experiencing a Dresselhaus effect from the ${\bf k}$-linear interaction.\cite{Tsitsishvili2004} Next, we explore the light-matter interactions of the spinor exciton via photoluminescence spectroscopy (PL).
As the time periods between absorption and emission of photons may range from short femtoseconds up to milliseconds, the microscopic description of the PL has to be accomplished by a semiconductor luminescence equations.\cite{Koch2006,Kira1998,Koch2006} Rather than tracing the dynamics of the optical transitions, finding the conservation laws supporting open channels for bright excitons in the electron-hole recombination is the first priority in the present work. An analysis of the magneto-PL spectrum of the spinor exciton can lead to an understanding
of how the presence of the SOIs influences the diamagnetic response\cite{Ambegaokar1990} observed
in spin degenerate excitons.
Furthermore, a thermodynamic persistent, charge current (CC), $I = -\partial F/ \partial \Phi$, has been theoretically suggested\cite{BUTTIKER1983365, Byers1961, Chakraborty1994} and was observed both in metal and semiconductor loops\cite{Levy1990, Chandrasekhar1991, Mailly1993, Jariwala2001} if the phase coherence of the charge wave functions throughout the whole sample is conserved. Similarly, the presence of SOIs provides the possibility to support persistent spin currents (SC).\cite{Sheng2006} Hence, without a loss of generality, we construct an ideal-ring model for (in)coherently-rotating excitons under the SOIs. Excitons are found to have a specific spin polarization, and to
possibly form a squeezed complex when the strength of the SOIs is adjusted. Whenever a coherent exciton is viewed as a neutral particle with an electric dipole moment, it acquires a nontrivial dual Aharonov-Casher (AC) phase\cite{Aharonov1984, Yi1995} when moving in the magnetic field. This implies the possibility to generate persistent dipole currents and spin dipole currents as well. The tunable SOIs open the possibility to coherently manipulate individual particles, and indirectly to manipulate the optical response of an exciton. By controlling the SOIs we can successfully realize a bright-dark sequence that
conceptually could be utilized as an element in an optical switch within the quantum regimes.

The paper is organized as follows. First, we introduce a theoretical model for an exciton in quantum rings subjected both to a magnetic flux and the SOIs. Then we analyze the AB energy spectra, magneto-optical response via the PL spectra, spin orientation, the generation of persistent currents, and the construction of the prototype of a quantum switch. Finally, we give a brief conclusion of the results.

%
%
%
%
%
\section{Hamiltonian}
The Hamiltonian of two-dimensional semiconductor quantum-ring excitons with Rashba and Dresselhaus spin-orbit interactions (SOIs) threaded by a magnetic flux perpendicular to the plane of the ring can be written as
\begin{equation}
      \hat{H_x} = \sum_{\substack{{\alpha = e,h}\\ i\,j}} H^0_{ij}(\alpha)\,\hat{c}^{\dag}_{\alpha\,i}\hat{c}_{\alpha\,j} + \frac{1}{2}\sum_{\substack{\alpha\beta=e,h\\ ii'jj'}} V_{ii'jj'}\, \hat{c}^{\dag}_{\alpha\,i}\hat{c}^{\dag}_{\beta\,j}\hat{c}_{\beta\,j'}\hat{c}_{\alpha\,i'},
\end{equation}
in which the sum is taken over band indices $(\alpha ,\beta)$ of the electron and hole bands $(e,h)$,
and quantum numbers $(i,j)$. Correspondingly, the field operator defined by $\hat{\Psi}({\boldsymbol r}) =\sum_{\substack{i}}\hat{c}_{i}\Psi_i({\boldsymbol r})$ is expressed as
\begin{align}
      \hat{\Psi}({\boldsymbol r}) &= \hat{\Psi}_e({\boldsymbol r}) + \hat{\Psi}^\dag_h({\boldsymbol r}) =\nonumber\\
      &\sum_{ k\,i, k>k_F} \hat{e}_{k\,i} \Psi_{k\,i}^e({\boldsymbol r}) + \sum_{k\,i, k<k_F} \hat{h}^\dag_{-k\,i} \Psi_{-k\,i}^{*h}({\boldsymbol r}),
\end{align}
where $k_F$ is the Fermi momentum, $\hat{e}$ and $\hat{h}$ are electron and hole annihilation operators, respectively, defined as $\hat{e}_{k,s} = \hat{c}_{k>k_F,s}$ and
$\hat{h}_{-k,-s}^+ = \hat{c}_{k<k_F,s}$, and
\begin{equation}
      \Psi^\alpha_{{k}\,i}({\boldsymbol r})= \psi_{i}({\boldsymbol r}_\alpha)u_{\alpha}({\boldsymbol k},{\boldsymbol r})
\end{equation}
is the Bloch function constructed by a two-component envelope function $\psi_{i}({\boldsymbol r}_\alpha)$ and the mutually orthogonal periodic function $u_{\alpha}({\boldsymbol k},{\boldsymbol r})$ of the conduction(valance) band in the unit cell.
In this manner, the evaluation of the single-particle energies for both an electron and a hole is equivalent to calculating the expectation values of the noninteracting Hamiltonian with the spinor wavefunctions given by $E_{\alpha\,i} = \langle \psi_{i}|H^0 | \psi_{i} \rangle_\alpha$. Typically, the Hamiltonian
$H^0 = K + V_{t} + Z_m + H_{D,R}$ includes
the kinetic energy from the canonical momentum $\Pi = {\boldsymbol p} \pm \frac{e}{c} {\boldsymbol A}$,
the trapping potential
\begin{equation}
      V_{t} = \frac{1}{2} m^* \omega {\boldsymbol r}^2 + \sum_{i=1,n_i}
      W_{i}\exp\left[{-\frac{({\boldsymbol r}-{\boldsymbol r}_{i})^2}{2\Delta_{i}}}\right],
\end{equation}
constructed by harmonic and shifted Gaussian wells,
the Zeeman interaction $Z_m = \frac{1}{2} g^* \mu_B B \sigma_z$, where $g^*$
represents the effective bulk Lande g-factor, and the Rashba and the Dresselhaus SOIs
\begin{equation}
      H_R = \frac{\lambda_R}{\hbar}\left[\Pi_y\sigma_x-\Pi_x\sigma_y\right]_h,
\end{equation}
and
\begin{equation}
      H_D = \frac{\lambda_D}{\hbar}\left[\Pi_x\sigma_x-\Pi_y\sigma_y\right]_e,
\end{equation}
respectively.

Since the presence of the SOIs couples the spin states, the application of a unitary transformation
via the operator
\begin{align}
      U =
      \exp\big[&-i\frac{m^*_h}{\hbar^2}\lambda_R(y_h\sigma_x-x_h\sigma_y)\nonumber\\
           &-i\frac{m^*_e}{\hbar^2}\lambda_D(x_e\sigma_x-y_e\sigma_y)\big],
\end{align}
diagonalizes the noninteracting system in the subspin spaces.\cite{ManuelValin-Rodriguez2002}
As a result, although ${L}_z$ and ${S}_z$ are coupled by the SOIs in a rotating reference frame,
the total angular momentum $J_z$ can be proved to commute with the Hamiltonian, \textit{i.e.},
$[J_z, \tilde{H}^0]_\alpha = 0$.
Therefore, the spinor wavefunctions in the rotating reference frame for both particles can be constructed by setting
\begin{equation}
      \psi^0({\boldsymbol r}_\alpha)
      = \left(%
      \begin{array}{c}
            \varphi^0_{l_\uparrow}(r_\alpha) e^{i l^\alpha_{\uparrow}\phi_\alpha} \\
            \varphi^0_{l_\downarrow}(r_\alpha) e^{i l^\alpha_{\downarrow} \phi_\alpha} \\
      \end{array}%
      \right), \mbox{ where } l^\alpha_{\downarrow}=l^\alpha_{\uparrow} + 1.
\end{equation}
Hence we can define the quasi-up and quasi-down wavefunctions for
both particles in the lab frame as
\begin{eqnarray}
\psi^u({\boldsymbol r}_e)&=&
\left(
  \begin{array}{c}
    1 \\
   -i\frac{m^*_e}{\hbar^2}\lambda_D r_e e^{-i\phi_e} \\
  \end{array}
\right) \varphi^0_{l_\uparrow}(r_e)e^{i l_\uparrow \phi_e},\label{equ-eu}\\
\psi^d({\boldsymbol r}_e)&=&
\left(
  \begin{array}{c}
    -i\frac{m^*_e}{\hbar^2}\lambda_D r_e e^{i\phi_e} \\
    1\\
  \end{array}
\right) \varphi^0_{l_\downarrow}(r_e)e^{i l_\downarrow \phi_e}, \label{equ-ed}
\end{eqnarray}
\begin{eqnarray}
\psi^u({\boldsymbol r}_h)&=&
\left(
  \begin{array}{c}
    1 \\
   -\frac{m^*_h}{\hbar^2}\lambda_R r_h e^{i\phi_h} \\
  \end{array}
\right) \varphi^0_{l_\uparrow}(r_h)e^{i l_\uparrow \phi_h}, \label{equ-hu}\\
\psi^d({\boldsymbol r}_h)&=&
\left(
  \begin{array}{c}
    \frac{m^*_h}{\hbar^2}\lambda_R r_h e^{-i\phi_h} \\
   1 \\
  \end{array}
\right) \varphi^0_{l_\downarrow}(r_h)e^{i l_\downarrow \phi_h},  \label{equ-hd}
\end{eqnarray}
and can construct an exciton wavefunction via a linear combination of electron-hole pair eigenfunctions
\begin{align}
      \Psi_x({\boldsymbol r}_e,{\boldsymbol r}_h)=\sum_{\mbox{\tiny$\begin{array}{c}
                                         \lambda_1=n_e,j_{z}(e) \\
                                         \lambda_2=n_h,j_{z}(h)
                                      \end{array}$}} a_{\lambda_1,\lambda_2,s_{z1},s_{z2}}\,
      \psi_{\lambda_1,s_{z1}}({\boldsymbol r}_e)\nonumber\\
      \otimes \psi_{\lambda_2,s_{z2}}({\boldsymbol r}_h).
\end{align}

While the connection between an electron-hole pair is described via Coulomb interaction, $V({\boldsymbol r}) = {-e^2}/{4\pi\epsilon_0 \epsilon|{\boldsymbol r}_e-{\boldsymbol r}_h|} \equiv {\beta_{eh}}/{|{\boldsymbol r}|}$,
it is more convenient to deal with the two-particle potential in the momentum representation by writing
\begin{eqnarray}
V({\boldsymbol r}) &=& \frac{1}{2\pi} \int V({\boldsymbol q}) e^{-i{\boldsymbol q}\cdot {\boldsymbol r}}\, d{\boldsymbol q}.
\end{eqnarray}
In this manner, a two-body function in the coordinate space such as $V({\bf r}) = 1/|{\bf r}_1 - {\bf r}_2|$, can be reduced to a single-body one in the momentum space by $V({\bf q}) = 2\pi \int\, J_0(|{\bf q}||{\bf r}_1 - {\bf r}_2|)\, d{\bf r} = 2\pi/|{\bf q}|$, in which $J_0$ is the first kind Bessel function of zeroth order.
Therefore, via Fourier transformation, and writing the single-particle wavefunction in terms of its Fourier conjugate
\begin{equation}
\psi ({\boldsymbol r}) =  \frac{1}{2\pi} \int \psi({\boldsymbol q}) e^{-i{\boldsymbol q}\cdot {\boldsymbol r}}\, d{\boldsymbol q},
\end{equation}
the calculation of Coulomb potential energy can be performed in the momentum space and leads to
\begin{equation}
V_{\lambda_1,\lambda_2,\lambda_3,\lambda_4} = \frac{1}{2\pi}\int\, C_{\lambda_1,\lambda_4}({\bf q})D_{\lambda_2,\lambda_3}({\bf q})V({\bf q})\, d{\bf q},
\end{equation}
in which $C_{\lambda,\lambda'}({\bf q}) = \int\, \psi_\lambda^*({\bf q}_1)\psi_{\lambda'}({\bf q}_1 + {\bf q})\, d{\bf q}_1$ and $D_{\lambda,\lambda'}({\bf q}) = \int\, \psi_\lambda^*({\bf q}_2)\psi_{\lambda'}({\bf q}_2 - {\bf q})\, d{\bf q}_2$ represent the momentum conservation of electron and hole with state quantum numbers $\lambda$ and $\lambda'$ during the potential scattering.
After analytically integrating the angular part and applying the Weber-Schafheitlin formula\cite{9789971506674} for Bessel function integrals, the evaluation of the matrix elements for the direct Coulomb interaction given by
\begin{widetext}
\begin{eqnarray}
V_{\lambda_1,\lambda_2,\lambda_3,\lambda_4} &=& \iint \psi^*_{\lambda_1,s_{z_1}} ({\boldsymbol r}_1) \psi^*_{\lambda_2,s_{z_2}} ({\boldsymbol r}_2)V({\boldsymbol r}_1-{\boldsymbol r}_2)  \psi_{\lambda_3,s_{z_3}} ({\boldsymbol r}_2) \psi_{\lambda_4,s_{z_4}} ({\boldsymbol r}_1)\, d{\boldsymbol r}_1\,d{\boldsymbol r}_2, \nonumber\\
& =& \delta_{{l}_e+{l}_h, {l}'_h+{l}'_e} \int_0^\infty q V(q) dq \int _0^\infty  \left[1+\left(\frac{m^*_e}{\hbar^2}\lambda_D r_e\right)^2\right]  J_{|{l}'_e-{l}_e|}(qr_e)  \nonumber\\
&& \times  \left[ \varphi_{n_e l_{e\uparrow}}^{*0}(r_e) \varphi_{n'_e l'_{e\uparrow}}^{0}(r_e)
  +  \varphi_{n_e l_{e\downarrow}}^{*0}(r_e) \varphi_{n'_e l'_{e\downarrow}}^{0}(r_e) \right] \,dr_e \nonumber\\
&& \times \int _0^\infty  \left[1+\left(\frac{m^*_h}{\hbar^2}\lambda_R r_h\right)^2\right]  J_{|{l}'_h-{l}_h|}(qr_h) \nonumber\\
&&\times \left[ \varphi_{n_h l_{h\uparrow}}^{*0}(r_h) \varphi_{n'_h l'_{h\uparrow}}^{0}(r_h)
  + \varphi_{n_h l_{h\downarrow}}^{*0}(r_h) \varphi_{n'_h l'_{h\downarrow}}^{0}(r_h) \right] \,dr_h
\label{eqn-v1234}
\end{eqnarray}
results in an analytical expression in terms of radial integrals
\begin{eqnarray}
V_{\lambda_1,\lambda_2,\lambda_3,\lambda_4} &=& \beta_{eh} \int _0^\infty  dr_e \int _0^{r_e} dr_h\, G_M\left(\frac{r_h^2}{r_e^2}\right) \nonumber\\
&&\times\frac{1}{r_e^{M+1}} \left[1+\left(\frac{m^*_e}{\hbar^2}\lambda_D r_e\right)^2\right]
\times \left[ \varphi_{n_e l_{e\uparrow}}^{*0}(r_e) \varphi_{n'_e l'_{e\uparrow}}^{0}(r_e)
  +  \varphi_{n_e l_{e\downarrow}}^{*0}(r_e) \varphi_{n'_e l'_{e\downarrow}}^{0}(r_e) \right] \nonumber\\
&&\times r^{M}_h \left[1+\left(\frac{m^*_h}{\hbar^2}\lambda_R r_h\right)^2\right]
\times   \left[ \varphi_{n_h l_{h\uparrow}}^{*0}(r_h) \varphi_{n'_h l'_{h\uparrow}}^{0}(r_h)
  +  \varphi_{n_h l_{h\downarrow}}^{*0}(r_h) \varphi_{n'_h l'_{h\downarrow}}^{0}(r_h) \right] \nonumber\\
&+& \beta_{eh}   \int _0^\infty  dr_h \int _0^{r_h}  dr_e\, G_M\left(\frac{r_e^2}{r_h^2}\right) \nonumber\\
&&\times\frac{1}{r_h^{M+1}} \left[1+\left(\frac{m^*_h}{\hbar^2}\lambda_R r_h\right)^2\right]
\times   \left[ \varphi_{n_h l_{h\uparrow}}^{*0}(r_h) \varphi_{n'_h l'_{h\uparrow}}^{0}(r_h)
  +  \varphi_{n_h l_{h\downarrow}}^{*0}(r_h) \varphi_{n'_h l'_{h\downarrow}}^{0}(r_h) \right] \nonumber\\
&&\times   r^{M}_e  \left[1+\left(\frac{m^*_e}{\hbar^2}\lambda_D r_e\right)^2\right]
\times \left[ \varphi_{n_e l_{e\uparrow}}^{*0}(r_e) \varphi_{n'_e l'_{e\uparrow}}^{0}(r_e)
  +  \varphi_{0,n_e l_{e\downarrow}}^{*0}(r_e) \varphi_{n'_e l'_{e\downarrow}}^{0}(r_e) \right],
\label{Vllll}
\end{eqnarray}
in which ${M}=|{l}_e-{l}'_e|=|{l}_h-{l}'_h|$ and
\begin{equation}
      G_M\left(\frac{r_2^2}{r_1^2} \right) = \frac{\Gamma\left(M+\frac{1}{2}\right)}{\Gamma(M+1)\Gamma\left(\frac{1}{2}\right)}
      \left[1 + \sum_{n=1}^\infty \frac{\Gamma\left(M+\frac{1}{2}+n\right)\Gamma\left(n+\frac{1}{2}\right)\Gamma\left(M+1\right)}
{n! \Gamma(M+1+n)\Gamma\left(\frac{1}{2}\right)\Gamma\left(M+\frac{1}{2}\right)} \left(\frac{r_2}{r_1} \right)^{2n}\right],
\end{equation}
\end{widetext}
is a convergent polynomial of the proper fraction $r_2/r_1$ with the Gamma function $\Gamma(x)$.
{We note that the expression for the matrix elements (\ref{Vllll}) is useful independent of
the exact radial confinement of the ring system.}
\subsection{Photoluminescence}
We next investigate the magneto-optical response of an exciton by exploring the photoluminescence spectrum. PL intensity is proportional to the transition rate $\Gamma$ that can be calculated via Fermi's golden rule regarding dipole-allowed light-matter interaction with the quantized electromagnetic fields (in terms of the annihilation and creation operators ${\hat{a}}_{\bf k}$ and ${\hat{a}}^\dag_{\bf k}$) of mode ${\bf k}$ and polarization $\epsilon_{\bf k}$. Within the
dressed-state representation, the rate of the photon emission process is given by
\begin{eqnarray} \label{e:emit}
      \Gamma &=& \frac{2\pi}{\hbar} |H_{if}'({\bf r})|^2 \delta(E_i-E_f) \nonumber\\
      &=& \frac{2\pi e^2}{\hbar} \left(\frac{\hbar \omega_{\bf k}}{2\epsilon V}\right) |\langle \beta, n_{\bf k} +1|{\hat{a}}^\dag_{\bf k}\epsilon_{\bf k}\cdot {\bf r} |\alpha, n_{\bf k}\rangle|^2 \nonumber\\
      &\ &\qquad\qquad\qquad \times\delta(E_\beta-E_\alpha +\hbar\omega_{\bf k})\nonumber\\
      &=& \frac{\pi e^2 \omega_{\bf k}}{\epsilon V}  (n_{\bf k}+1)|\langle \beta|\epsilon_{\bf k}\cdot {\bf r} |\alpha \rangle|^2\nonumber\\
      &\ &\qquad\qquad\qquad \times\delta(E_\beta-E_\alpha +\hbar\omega_{\bf k}),
\end{eqnarray}
in which the term that is (not) proportion to $n_{\bf k}$ accounts for the stimulated (spontaneous) emission.
To calculate the total spontaneous emission rate, we need to sum over all photon
modes ${\bf k}$ and choose to use the continuous representation
\begin{equation}
   \sum_{\bf k} = \frac{1}{\Delta{\bf k}} \sum_{\bf k} \Delta{\bf k} \approx \frac{V}{8\pi^3} \int d{\bf k}.
\end{equation}
Then we obtain
\begin{align}
      \Gamma_{i \rightarrow f} = \frac{e^2\eta^2\omega^3_{\bf k}}{8\pi^2 \hbar\epsilon c^3}
      \int_0^\infty \int_0^\pi \int_{0}^{2\pi} |\langle \beta|\epsilon_{\bf k}\cdot {\bf r}|\alpha\rangle|^2 \nonumber\\
      \delta(\omega_{\beta\alpha}+\omega_{\bf k})d\omega_{\bf k} \sin\theta_k d\theta_k d\phi_k,  \label{eqn-Trate}
\end{align}
in which $\eta$ is the refractive coefficient of the material.

We evaluate Eq.\ (\ref{eqn-Trate}) by
rotating the reference frame such that the dipole vector $\boldsymbol{r}$ points in the $x$ direction,
labeled by ${\boldsymbol r} = (r, 0, 0)$, and the wave vector of the laser pulse is $(k\sin\theta_k\cos\phi_k, k\sin\theta_k\sin\phi_k, k\cos\theta_k)$.
To satisfy the orthogonal criteria, one field polarization vector $\boldsymbol{\epsilon}_{{\boldsymbol k}1}$
can be chosen to lie in the ${k}$-${z}$ plane having coordinate components
$\boldsymbol{\epsilon}_{{\bf k}1}
= (
                          \begin{array}{ccc}
                            -\cos\theta_k\cos\phi_k,& -\cos\theta_k\sin\phi_k,& \sin\theta_k \\
                          \end{array}
                        )$,
hence
the second one takes the form
$\boldsymbol{\epsilon}_{{\boldsymbol k}2} = {\boldsymbol k}\times \boldsymbol{\epsilon}_{{\boldsymbol k}1}
= \sin\phi_k~\hat{\boldsymbol{i}} - \cos\phi_k~\hat{\boldsymbol{j}}$.
As a consequence, the PL intensity of the dipole transition results in
\begin{eqnarray}
  \Gamma_{i \rightarrow f}
  = \frac{e^2\eta^3\omega_{\alpha\beta}}{3\pi \hbar\epsilon c^3} |{\bf r}_{\alpha\beta}|^2.
  \label{Eqn-pl_rate}
\end{eqnarray}

Evaluation of Eq.~(\ref{Eqn-pl_rate}) in semiconductor nanostructures can be carried out by defining the one-particle polarization operator ${\hat P}$ given by
\begin{equation}
      {\hat P} =\sum_{\mu,\nu=e,h} \int d{\bf r}\,\hat{\Psi}_\mu^\dag({\bf r})\, e\,{\bf r}\, \hat{\Psi}_\nu({\bf r}).
\end{equation}
In what follows we only consider direct optical transitions. We
apply a $k\cdot p$ approximation and expand
$k$ around the band bottom to simulate a strong recombination
that occurs at the $\Gamma$-point in III-V heterostructures.
Eventually, the polarization operator for annihilating an electron-hole pair is written as
\begin{align}\label{e:peh}
      {\hat P}_{eh}
      \approx  \sum_{\lambda,\lambda'} \hat{e}_\lambda \hat{h}_{\lambda'}
      &\int d{\bf R}\,\psi^{e}_{\lambda}({\bf R}) \psi^{h}_{\lambda'}({\bf R}) \nonumber\\
      &\int_{\substack{unit\\ cell}} d{\bf r}\,e{\bf r}\,u_c({\bf r})u_v({\bf r}).
\end{align}
However, the applied magnetic flux
causes a cross effect with the time-dependent electromagnetic fields. Let ${\bf A}^0$ be the static field, the external perturbation involves
${\bf A}({\bf r},t)\cdot ({\bf p}+ e{\bf A}^0) $, which means, while
$\hat{P}_{eh} = \sum_{\lambda,\lambda'} P^{eh}_{\lambda,\lambda'}
{\hat{e}}_\lambda {\hat{h}}_{\lambda'}$,
the matrix element of the annihilation of an electron-hole pair should be constructed via
$\langle 0|\hat{H'}|eh \rangle_{nm}$, which is proportional to $\sum_{\lambda,\lambda'} P^{eh}_{\lambda,\lambda'} \langle 0| {\hat e}_\lambda {\hat h}_{\lambda'}|e_n h_m \rangle$,
in which $\langle 0|$ denotes the vacuum state with exciton vacancies. Therefore, in the presence of crossed external fields, the matrix element of the dipole transition takes the form
\begin{eqnarray}
      M_{\lambda,\lambda'} &=& \int u_c({\bf r}) \psi_\lambda^{e} \bigg[\frac{e}{m} {\bf A}({\bf r},t)\cdot ({\bf p}+ e {\bf A}^0)\bigg] u_v({\bf r}) \psi_{\lambda'}^{h}\, d{\bf r} \nonumber \\
      &=& \sum_{Ncell}  \psi_\lambda^{e} \psi_{\lambda'}^{h} \int_{\substack{unit \\cell}} u_c({\bf r})
      \big[{\bf A}({\bf r},t)\cdot {\bf p}\big] u_v({\bf r}) d{\bf r} \nonumber\\
      &&+  \sum_{Ncell}  \bigg[\psi_\lambda^{e}({\bf A}({\bf r},t)\cdot {\bf p})\psi_\lambda^{h} +  \psi_\lambda^{e}({\bf A}({\bf r},t)\cdot e{\bf A}^0)\psi_{\lambda'}^{h}\bigg] \nonumber\\
      && \qquad\qquad \times\int_{\substack{unit \\cell}} u_c({\bf r})u_v({\bf r}) \, d{\bf r} \nonumber\\
      &=& \int \,\psi_\lambda^{e}({\bf R}) \psi_{\lambda'}^{h}({\bf R}) d{\bf R} \nonumber\\
      && \qquad\qquad \times\int_{\substack{unit \\cell}} u_c({\bf r}) [{\bf A}({\bf r},t)\cdot {\bf p}]u_v({\bf r}) d{\bf r}. \label{Eqn-Mvc}
\end{eqnarray}
While the velocity gauge field ${\bf A}\cdot {\bf p}$ can be equivalently converted to ${\bf E}\cdot {\bf r}$ in the length gauge and the integration of the periodic functions over the unit cell can be dealt with as in the bulk, Eq.~(\ref{Eqn-Mvc}) has been proven to be adequate for the calculation of the transition rate for a spontaneous recombination.

Extensive exploration of the SOI associated excitonic PL spectrum is made feasible
by replacing the spin degenerate charge wavefunctions in Eq.~(\ref{Eqn-Mvc}) with the spinors
of Eqs.~(\ref{equ-eu}-\ref{equ-hd}).
While the recombination destroys the excitonic states,
the energy is compensated in the way of photon radiation
whenever the particles in the electron-hole pair have opposite rotational parity
and spin orientations.
In other words, the effective recombination of bright excitons is associated with
\begin{eqnarray}
      R_{eh}
      &=& \int rdr d\phi\, \Big[\sum_{l_\uparrow,l'_\downarrow} c_{l_\uparrow,l'_\downarrow}
      \varphi^0_{l_\uparrow}(e)\varphi^0_{l'_\downarrow}(h)
      e^{i (l_\uparrow + l'_\downarrow) \phi}\nonumber\\
      && \qquad +
      \sum_{l_\downarrow,l'_\uparrow} d_{l_\downarrow,l'_\uparrow}
      \varphi^0_{l_\downarrow}(e)\varphi^0_{l'_\uparrow}(h)
      e^{i (l_\downarrow + l'_\uparrow) \phi}
      \Big].
\label{eqn-Fvc}
\end{eqnarray}
Substituting Eq.\ (\ref{eqn-Fvc}) into the dipole transition of
Eq.\ (\ref{Eqn-pl_rate}), we find that for excitonic recombination between
$|\alpha\rangle$ and $|\beta\rangle$ states, $I(\alpha,\beta) \propto |R_{eh}|^2$.
At finite temperatures, there is a considerable possibility to detect photon emissions from high-level
recombination and therefore the total PL intensity $I_T$
that sums up contributions from states $\{|n\rangle\}$ with energies $\{E_n\}$ by
the classical Boltzmann distribution can be expressed as
\beq
    I_T = {\displaystyle\sum_{n=1}^\infty I_n e^{-E_n/k_B T}}{\Big /}{\displaystyle\sum_{n=1}^\infty e^{-E_n/k_B T}}.
\label{eqn:IT}
\eeq

\section{Persistent Currents}

The charge current density $\textit{\textbf{j}}_c$ is the expectation value of
the charge current operator $\hat{\boldsymbol{j}}_c=
[\hat{{\rho}}(\boldsymbol{r})\hat{\boldsymbol{v}} + \hat{\boldsymbol{v}}\hat{{\rho}}(\boldsymbol{r})]/2$,
where $\hat{\rho}(\boldsymbol{r}) = q \delta(\boldsymbol{r}-\boldsymbol{r}^\prime)$ is the charge density operator.
Eventually, in matrix representation, it
can be written as
$\boldsymbol{j}_c (\boldsymbol{r}) = q\,\mathrm{Re}\left\{\psi^\dag(\boldsymbol{r}) \hat{\boldsymbol{v}}\,\psi(\boldsymbol{r}) \right\}$.
Similarly, the spin density operator defined by
$\hat{\boldsymbol{s}}(\boldsymbol{r}) = \hbar\hat{\boldsymbol{\sigma}}\delta(\boldsymbol{r}-\boldsymbol{r}')/2$,
where $\hat{\boldsymbol{\sigma}}$ denotes Pauli matrices, leads to
the spin current density operator
$\hat{\boldsymbol{j}}_s(\boldsymbol{r}) = [\hat{\boldsymbol{s}}(\boldsymbol{r})\hat{\boldsymbol{v}} +
\hat{\boldsymbol{v}}\hat{\boldsymbol{s}}(\boldsymbol{r})]/2$ and
%
gives the spin current density
$\boldsymbol{j}_s (\boldsymbol{r}) = \mathrm{Re}\left\{\psi^\dag(\boldsymbol{r})\, \hat{\boldsymbol{v}}\hat{\boldsymbol{s}}(\boldsymbol{r})\,\psi(\boldsymbol{r}) \right\}$.
%
Since the velocity of a particle of a corresponding Hamiltonian $H$ is given by
$\boldsymbol{v} = \hbar^{-1} \partial H/\partial \boldsymbol{k} = \partial H/\partial \boldsymbol{p}$, for an electron in an ideal 1D ring threaded by a perpendicular magnetic flux $\Phi$ under the influence of SOIs, it moves along the circumference of the ring with the velocity
\begin{equation}
\boldsymbol{v}_\phi = \frac{1}{m^*}\left(\boldsymbol{P} + \frac{e}{c} \boldsymbol{A}\right)_\phi
       + \lambda_R \sigma_r
       - \lambda_D \sigma_\phi(-\phi).
\end{equation}
%
{This implies that the current densities can be evaluated from the mean values of the Hamiltonian eigenstates $|n\rangle$. Then after some arithmetic one obtains}
the persistent charge current
\begin{equation}
I_{c,n} = \frac{e}{2\pi \hbar}  \frac{\partial E_n}{\partial f}, \label{Icn}
\end{equation}
together with the spin currents
\begin{equation}
I_{s,n}^{s_z} = \frac{1}{2\pi \hbar} \left(\frac{\partial E_n}{\partial f}\right)\,\sigma\cos\Theta, \label{Isn}
\end{equation}
in the presence of the SOIs, where $f = \Phi/\Phi_0$, with $\Phi_0 = hc/e$ as the flux quantum$, \sigma = \pm 1$ denoting the spin-up(-down) state, respectively and $\Theta$ is the angle between the spin vector and the z-axis.

When assuming ballistic collisions in an ideal ring, the thermal equilibrium properties as well as the generation of persistent currents of the spinor excitons can be investigated analytically.
In a hard-ring 1D confinement, an electron has the energy
\begin{align}
      E_{\uparrow,\downarrow} = \frac{\hbar^2}{2m_e^*R_e^2}(l_{\uparrow,\downarrow} + f)^2 &\pm \frac{m_e^*}{\hbar^2}{\lambda}^2_D
      (l_{\uparrow,\downarrow}+f)\nonumber\\
      &\pm \frac{m_e^*}{\hbar^2}{\lambda}^2_D (f\mp 1). \label{eqn:Ee-1d}
\end{align}
Eqs.~(\ref{Icn}) and (\ref{Isn}) lead to the charge current and the spin current being expressed as
\begin{eqnarray}
      \label{Ie}
      I_{l,\sigma} &=& - \frac{q}{2\pi\hbar}\left[\frac{\hbar^2}{m_e^*R_e^2}(l + f)+ 2\frac{m_e^*}{\hbar^2}{\lambda}_D^2\,\sigma \right], \\
      I_{l,\sigma}^{s_z} &=& \frac{1}{2\pi\hbar} \left[\frac{\hbar^2}{m_e^*R_e^2}(l + f)+ 2\frac{m_e^*}{\hbar^2}{\lambda}_D^2\,\sigma \right]\sigma
      \cos\Theta_e. \nonumber
\end{eqnarray}
Similarly, the eigenenergy of a hole can be written as
\begin{align}
      E_{\uparrow,\downarrow} = \frac{\hbar^2}{2m_h^*R_h^2}(m_{\uparrow,\downarrow}-f)^2
      &\pm \frac{m_h^*}{\hbar^2}{\lambda}_R^2 (f- m_{\uparrow,\downarrow})\nonumber\\
      &\pm \frac{m_h^*}{\hbar^2}{\lambda}_R^2 (f \mp 1). \label{eqn:Eh-1d}
\end{align}
Correspondingly, the persistent currents are
\begin{eqnarray}
      \label{Ih}
      I_{m,\sigma}&=& \frac{q}{2\pi\hbar}\left[\frac{\hbar^2}{m_h^*R_h^2}(f-m)+ 2\frac{m_h^*}{\hbar^2}{\lambda}_R^2\,\sigma\right], \\
      I_{m,\sigma}^{s_z} &=& \frac{q}{2\pi\hbar} \left[\frac{\hbar^2}{m_h^*R_h^2}(f-m) + 2\frac{m_h^*}{\hbar^2}{\lambda}_R^2\,\sigma \right]\sigma
      \cos\Theta_h. \nonumber
\end{eqnarray}

To find the magnitudes of $\Theta_{e(h)}$, we need to examine the spin orientation $\langle \boldsymbol{S}\rangle$ of single particles. While writing
$\psi^u({\bf r}_e)=
(
  \begin{array}{cc}
    \cos\theta & -i\sin\theta e^{-i\phi_e}
  \end{array}
)^T e^{i l_\uparrow \phi_e}$
%
%
and $\psi^d({\bf r}_e)=
(
  \begin{array}{cc}
    -i\sin\theta e^{i\phi_e} & \cos\theta
  \end{array}
)^T e^{i l_\downarrow \phi_e}$,
we obtain
\begin{align}
      \langle S^e_{\uparrow,\downarrow}\rangle
      = \mp \frac{\hbar}{2}\big[\sin 2\theta (\sin\phi_e \,\hat{x}
      +& \cos\phi_e \,\hat{y})\nonumber\\
      &- \cos 2\theta \,\hat{z}\big],
      \label{Seud}
\end{align}
and find that $\Theta_e=2\theta$.
Similarly, by writing $\psi^u({\bf r}_h)=
(
  \begin{array}{cc}
    \cos\eta \, &  \sin\eta e^{i\phi_h}
  \end{array}
)^T e^{i l_\uparrow \phi_h}$ and
$\psi^d({\bf r}_h)=
(
  \begin{array}{cc}
    \sin\eta e^{-i\phi_h} & \cos\eta
  \end{array}
)^T e^{i l_\downarrow \phi_h}$,
we obtain
\begin{align}
      \langle S^h_{\uparrow,\downarrow}\rangle
      = \mp \frac{\hbar}{2}\big[\sin 2\eta (\cos\phi_h \,\hat{x}
      +& \sin\phi_h \,\hat{y})\nonumber\\
      &- \cos 2\eta \,\hat{z}\big], \label{Shud}
\end{align}
and find that $\Theta_h=2\eta$.
The two-particle exciton system has to be investigated in a four dimensional space. Spanned by the four-component basis
$\{|++\rangle,\, |+-\rangle,\, |-+\rangle,\, |--\rangle\}$, the spin orientation of the up(down)-pair and mixed-pair exciton is represented by
\begin{align}
      \langle S_{\uparrow\uparrow,\downarrow\downarrow}\rangle = \mp
      \frac{\hbar}{2} \Big[\big(&\sin 2\eta\cos\phi_h +\sin 2\theta \sin\phi_e\big)\,\hat{x}\nonumber\\
      &+\big(\sin 2\eta\sin\phi_h + \sin 2\theta\cos\phi_e\big) \,\hat{y} \nonumber\\
      &- 2\cos(\theta+\eta)\cos(\theta-\eta)\,\hat{z} \Big]. \label{Suu}
\end{align}
\begin{align}
    \langle S_{\uparrow\downarrow,\downarrow\uparrow}\rangle = \pm
    \frac{\hbar}{2} \Big[\big(&\sin 2\eta\cos\phi_h -\sin 2\theta \sin\phi_e\big)\,\hat{x}\nonumber\\
    &+ \big(\sin 2\eta\sin\phi_h -\sin 2\theta\cos\phi_e\big) \,\hat{y} \nonumber\\
    &- 2\sin(\theta+\eta)\sin(\theta-\eta)\,\hat{z} \Big]. \label{Sud}
\end{align}
Eqs.~(\ref{Seud}-\ref{Sud}) reveal that the single-particle picture is approximately
valid to determine the SC of an incoherent exciton, but below we show that for the coherent case, the SC is
related to the resultant of the two-particle spin vectors.

     A coherent exciton can be viewed as a neutral
particle with an electric dipole moment ${\bf d}$; thus it is also referred to as a dipolar exciton.
Moving in the magnetic field ${\bf B}$, the exciton acquires a nontrivial dual Aharnov-Casher phase defined by
\begin{equation}
      S_{dAC} = \oint ({\boldsymbol B}\times {\boldsymbol d})\cdot d{\boldsymbol l} = \frac{e}{\hbar c}\Phi_d.
\end{equation}
As a result, an equilibrium persistent dipole current can be defined by
\begin{equation}
      I_{d} \equiv  \frac{d}{2\pi \hbar} \sum_n \frac{\partial E_n}{\partial f_d}.
\end{equation}
For simplicity, setting $f_d = \Delta \Phi/\Phi_0 = (\Phi_h-\Phi_e)/\Phi_0$, a dipole can be viewed as a complex with mass $m_d = (m_e^* R_e^2 + m_h^* R_h^2)/R_d^2$, where $R_d = (R_e + R_h)/2$ is set as an effective length threaded by the magnetic field.

Coherent excitons can be classified into two classes. The first class consists of full up(down)-pair particles and their energies are represented by
\begin{align}
    E^{U,D}_x = \frac{\hbar^2}{2m_d R_d^2}(L-f_d)^2 &+ \frac{m_d}{\hbar^2}{\lambda}_+^2 \big|(2f_d-L)\big|\,\sigma \nonumber\\
    &+ \frac{1}{2}g_{ex} f_d\,\sigma,
\end{align}
whereas the second class consists of excitons with mixed-pair particles having energies
\begin{align}
    E^{u,d}_x =\frac{\hbar^2}{2m_d R_d^2}(L-f_d)^2 &+ \frac{m_d}{\hbar^2}{\lambda}_-^2 \big|(2f_d-L)\big|\,\sigma \nonumber\\
    &+ \frac{1}{2}{g}_{ex} f_d\,\sigma.
\end{align}
In both cases, $L$ represents the total angular momentum of the exciton,
${\lambda}^2_+ = {\lambda}^2_D + {\lambda}^2_R$, ${\lambda}^2_- = |{\lambda}^2_D - {\lambda}^2_R|$, and ${g}_{ex}$
{are used to represent the effective SOIs and Lande g-factors of the exciton.}

The dipole currents (DC) and spin dipole currents (SDC) for effective spin up$(\uparrow)$ and down$(\downarrow)$ excitons can then be expressed in an organized way as
\begin{align}
    I^{\uparrow\downarrow}_{d\,L}= \frac{d}{2\pi\hbar}\big[\frac{\hbar^2}{m_d R_d^2}&(f_d-L)\nonumber\\
    &+\Big(2 \frac{m_d}{\hbar^2}{\lambda}_{\pm}^2
    + \frac{{g}_{ex}}{2}\big)\,\sigma \Big], \label{eqn:I-DC}
\end{align}
and
\begin{align}
    I^{s_z,\uparrow\downarrow}_{d\,L}= \frac{1}{2\pi\hbar}\Big[\frac{\hbar^2}{m_d R_d^2}&(f_d-L)\nonumber\\
    &+\Big(2 \frac{m_d}{\hbar^2}{\lambda}_{\pm}^2
    + \frac{{g}_{ex}}{2}\Big)\,\sigma \Big]\sigma\,p^\pm, \label{eqn:I-SDC}
\end{align}
in which
$p^+ = \cos({\theta+\eta})\cos({\theta-\eta})$
and
$p^- = \cos({\theta+\eta+\pi/2})\cos({\theta-\eta-\pi/2})$.

%

\section{Discussion}

By considering the InAs
heterostructures, all calculations in this work are implemented and represented dimensionlessly based on the unit scales: the effective mass for the electron and the heavy hole $m^*_e = 0.026\, m_e$ and $m^*_h = 0.33\,m_e$, respectively, where $m_e$ is the free electron mass; the length $l^* = \sqrt{(\hbar/m^*_e \omega)} = 2\times 10^{-8}$\,m and the energy $E^* = \hbar\omega$; the static dielectric constant $\epsilon = 14.6$ and the single-particle Lande $g$-factors $g_e = -0.6$ and $g_h = -1.6$. For the two-particle system, the dimensionless scalings of kinematic parameters are performed in reference to the electron's coordinates. For SOIs, the unit strength is given by $\lambda = 10^{-11}$\,eV-m, and the strength parameter is abbreviated in terms of its magnitude $\lambda_{D,R}$.
Using the measurement for the bulk, the energy gap is set to $E_g = 430$\,meV. For convenience, the common constant $E_g$ is extracted from the calculation of total energies of the electron-hole pair and the exciton.

\subsection{Magneto-Optical Spectra}
As a magnetic flux threads the orbits of the electron and the hole in a quantum ring, the two particles acquire opposite geometric phases $e^{\pm ief(r_{e,h})/hc}$ when completing a full circulation.
Assuming that the electron and the hole move independently in opposite directions, a phase factor $\cos^2(\Phi/2\Phi_0)$, in which $\Phi=-[f_{e,h}(2\pi)-f_{e,h}(0)]$ extracted from the probability distribution of finding an individual particle provides the evidence of the interference effect.
As a result, for both a noninteracting electron-hole pair and an exciton in a quantum ring, the interference patterns reveal that for low-lying states, the energies associated with the corresponding single-particle angular momenta vary with the period of the flux quantum.
However, the coexistent antiparallel circulation leads to an effectively nonrotating ground state, 
turning the compound particle into a scalar exciton, eliminating the AB feature.

On the other hand, if there is a net flux traversing the electron and the hole orbits,
i.e., $|\Phi_h-\Phi_e| = \Delta \Phi > 0$, the two particles become correlated and can be regarded as a whole.
In this case, the charges are moving coherently carrying a common phase as the complex completes a full circulation.
As a consequence, with or without the SOIs, the ground state spectra demonstrate an AB effect.

Whenever the SOIs tend to flip a spin orientation, the down-spin ground state
causes a blue shift of the AB cross-over. Considering an ideal-ring model, 
the cross-over occurs at $\Phi/\Phi_0 = (2n+1)/2,\, n=0,\,1,\,2,\,\ldots$ in the absence of the SOIs. However, the introduction
of the additional contribution $\lambda_{D,R}$ to the energy gives rise to an energy blue-shift in the presence of the SOIs.
Therefore, in realistic rings where the effective range of the SOIs becomes spatially dependent, a pronounced shift should be observed.

\begin{figure}[htbp!]
\includegraphics[width=0.45\textwidth,bb=0 0 596 681,clip]{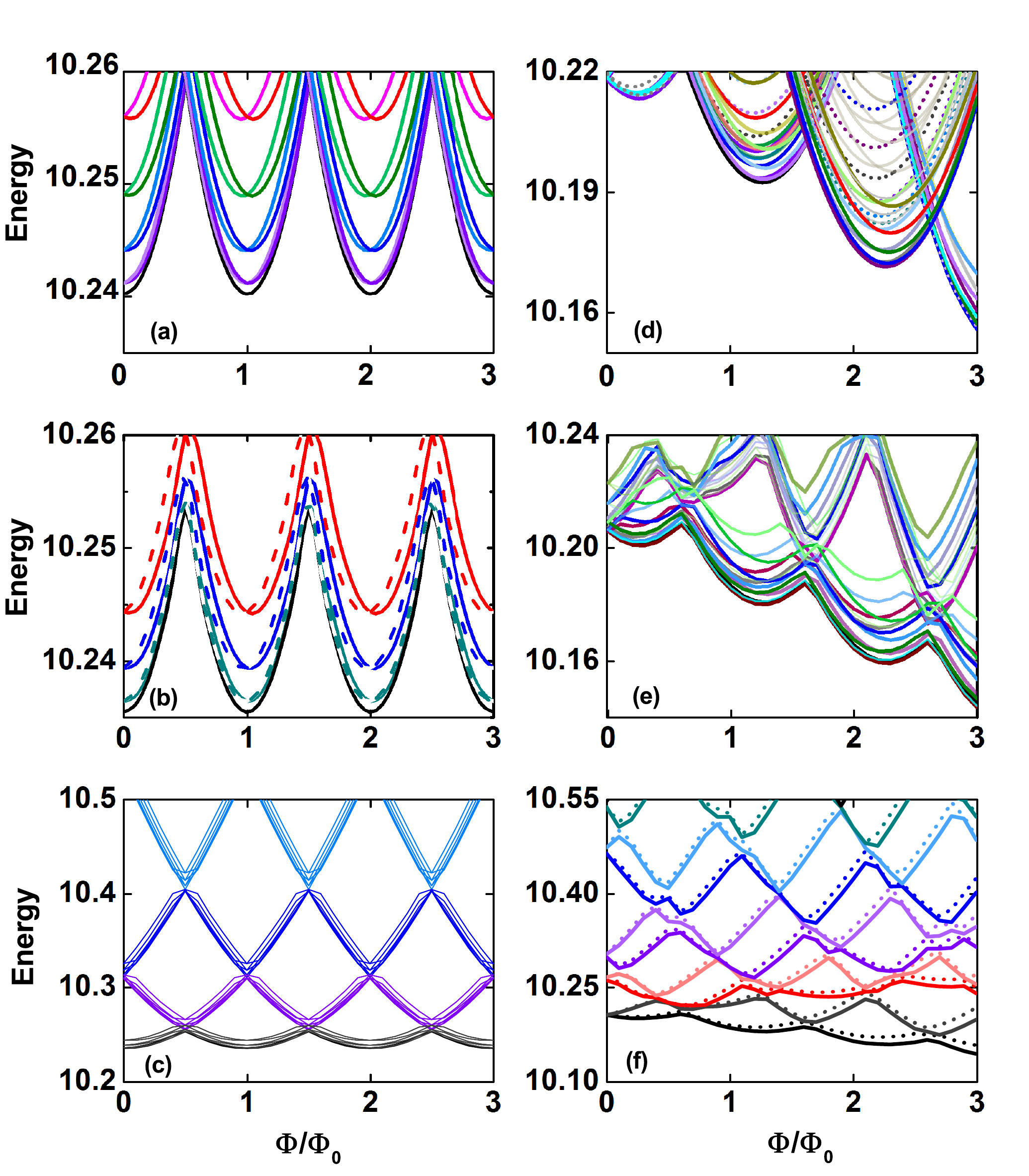}
\caption{(Color online)
Energy spectra of the scalar exciton in the absence (left panel) and presence (right panel) of SOIs. (a) and (d) demonstrate the energy spectra for noninteracting electron-hole pair and for the former they are corresponding to $L = 0(\mathrm{black}),\, \pm 1(\mathrm{purple}),\, \pm 2(\mathrm{blue}),\, \pm 3(\mathrm{green})$ and $\pm 4(\mathrm{red})$ phases, whereas for the latter the ground state shifts to $L = -1, J =-2(\mathrm{purple})$ and the excited states obeying different rules in changing the hole's angular momentum are split into subgroups labeled with dark, light, and dashed lines. The mutual-particle interaction is considered for the energy spectra for an exciton shown in (b) and (e), in which the system is specified by the exciton angular momentum determined according to the selection rules. With a large scale view of the spectra, (c) and (f) reveal the optical interference in the cluster of low-lying energy levels and in the bright exciton states, respectively.
}
\label{Fig1}
\end{figure}

\begin{figure}[thbp!]
\includegraphics[width=0.48\textwidth,bb=0 0 614 429,clip]{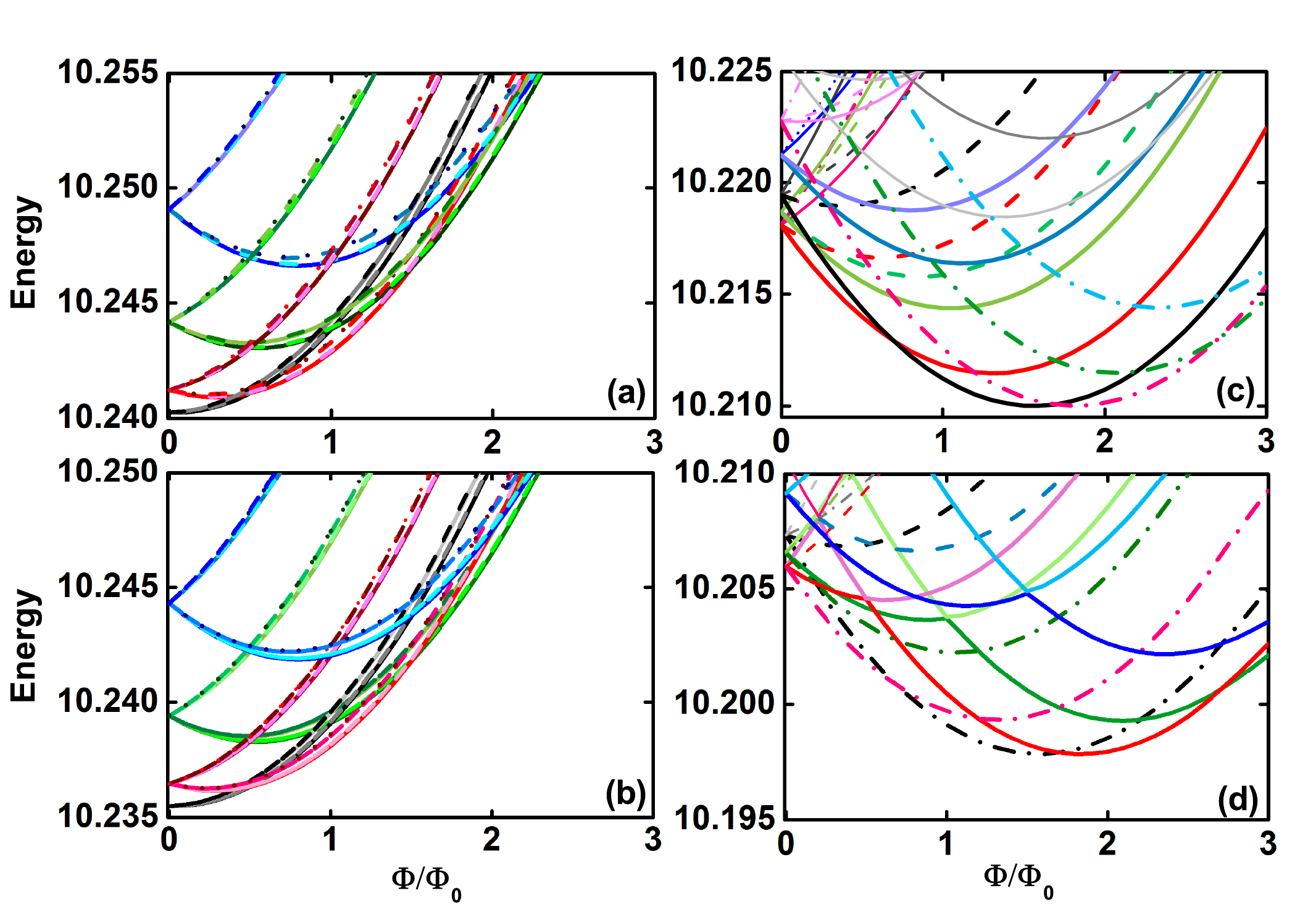}
\caption{(Color online)
Energy spectra of the dipolar exciton in the absence (left panel) and presence (right panel) of the SOIs. The upper figures demonstrate the low-lying energy spectra for noninteracting electron-hole pairs.  In (a), the lower[upper] branches belong to the spinor states with $L_e = 0$ and $L_h = 0(\mathrm{black}), 1[-1](\mathrm{red}), 2[-2](\mathrm{green}), 3[-3](\mathrm{blue})$. In (c), the ordering of the ground state in $L = -1; J = -2$({red}), $L = 0; J = -1$({black}), $L = 1; J = 0$({dashed-dot pink}), $L = 2; J = 1$({dashed-dot green}), $L = 3; J = 2$({dashed-dot blue}) demonstrates the AB feature. The lower figures show the energy spectra for the exciton.
In (b), the lower[upper] branches belong to the state families with total orbital angular momenta $L = 0(\mathrm{black}), 1[-1](\mathrm{red}), 2[-2](\mathrm{green}), 3[-3](\mathrm{blue})$. In (d), the ordering of the ground state for the spinor families with $L = -1$(dashed-dot pink), $L = 0$(dashed-dot black), $L = 1$(red), $L = 2$(green), and $L = 3$(blue) is shown.
}
\label{Fig2}
\end{figure}

  Fig.~\ref{Fig1}(a-c) and \ref{Fig2}(a-b) show that
without SOIs, the flux-dependent two-particle ground states all belong to the phase with zero angular momenta $L = 0$ and $J = 0,\,\pm 1$. Phases of bottom-up orders are labeled by $L = 0(\mathrm{black}),\, \pm 1(\mathrm{purple}),\, \pm 2(\mathrm{blue}),\, \pm 3(\mathrm{green})$ and $\pm 4(\mathrm{red})$, in which the left-to-right subphases are specified according to the rule
$\{l_e\,\uparrow\downarrow, l^{n-1}_h \pm 1\,\uparrow\downarrow\}$.
For example,
$1^{st}$ level $\supseteq$\,$\{$$\{0\,\uparrow\downarrow,0\,\uparrow\downarrow\}$, $\{-1\,\uparrow\downarrow,1\,\uparrow\downarrow\}$,
$\{-2\,\uparrow\downarrow,2\,\uparrow\downarrow\}$,
$\{-3\,\uparrow\downarrow,3\,\uparrow\downarrow\}$$\}$ and
$2^{nd}$ level $\supseteq$\,$\{$$\{0\,\uparrow\downarrow,\pm 1\,\uparrow\downarrow\}$, $\{-1\,\uparrow\downarrow,0,2\,\uparrow\downarrow\}$,
$\{-2\,\uparrow\downarrow,1,3\,\uparrow\downarrow\}$,
$\{-3\,\uparrow\downarrow,2,4\,\uparrow\downarrow\}$$\}$.

On the other hand, a neutral system prefers to occupy the $L = -1, J =-2(\mathrm{purple})$ state in
the presence of the SOIs. As the mutual-particle interaction is considered, the system is specified by the exciton angular momentum determined according to the selection rules.
The Coulomb interaction (Eq.~(\ref{eqn-v1234})) causes occupation of states with differing definite orbital and spin angular momenta and the total angular momenta as well. As SOI is involved, the spin flip is allowed, and therefore those states which fulfill the conservation rules, the conservation of angular momentum in the scattering, $L_e + L_h = L_e' + L_h'$ and $S_e + S_h = S_e' + S_h'$, make contributions to the energy spectrum.
As a consequence, the inclusion of Coulomb interaction results in a complicated crossed pattern for the spectra as shown in Fig.~\ref{Fig1}(b) and (e). While the periodic oscillating feature is left untouched only for the first few low-lying states, the excited states obeying different rules in changing the hole's angular momentum are split into subgroups labeled with dark, light, and dashed lines. For example, the first subgroup states consist of the $J=\pm 1$ plus the $\pm 2$ degenerate states that mainly contribute to the $J_e = \pm 1/2$ degeneracy of an electron and the $J_h = \pm 3/2$ degeneracy of a hole. All these states demonstrate a 4-fold degeneracy at $B = 0$ and reveal that the time-reversal symmetry is still preserved even under the Coulomb field.
With a large scale view of the spectra, (c) and (f) reveal the optical interference in the cluster of low-lying energy levels and in the bright exciton states, respectively.

The energy spectra of a coherently moving exciton are shown
in Fig.~\ref{Fig2}. The left (right) panel displays the spectra without (with) the inclusion of SOIs. The upper (lower) figures demonstrate the low-lying energy spectra for noninteracting electron-hole pairs (excitons).
As mentioned above, the AB features are clearly revealed in the ground state. Without the SOIs,
the magnetic field splits the energy levels into two branches with near spin degeneracy.
In (a), the lower[upper] branches belong to the spinor states with $L_e = 0$ and $L_h = 0(\mathrm{black}), 1[-1](\mathrm{red}), 2[-2](\mathrm{green}), 3[-3](\mathrm{blue})$.
In (b), the lower[upper] branches belong to the state families with total orbital angular momenta $L = 0(\mathrm{black}), 1[-1](\mathrm{red}), 2[-2](\mathrm{green}), 3[-3](\mathrm{blue})$.
It is worth noticing that, from the Hamiltonian, the SOIs tend to polarize both the electron and the hole into the down-spin phase, initiating a periodic transfer from the $L = -1$ rather than the $L = 0$ state at $B = 0$. Furthermore, the SOIs cause a fine splitting manifold but leave the Kramer's degeneracy intact until an external $B$ field is applied. In (c), the ground state shows its ordering in $L = -1; J = -2$({red}), $L = 0; J = -1$({black}), $L = 1; J = 0$({dashed-dot pink}), $L = 2; J = 1$({dashed-dot green}), $L = 3; J = 2$({dashed-dot blue}). In (d), the ground states for the spinor families with $L = -1$(dashed-dot pink), $L = 0$(dashed-dot black), $L = 1$(red), $L = 2$(green), and $L = 3$(blue) are shown.
%

%
%
%
%

\subsection{Photoluminescence Spectra}
\begin{figure}[thbp!]
\includegraphics[width=0.48\textwidth,clip]{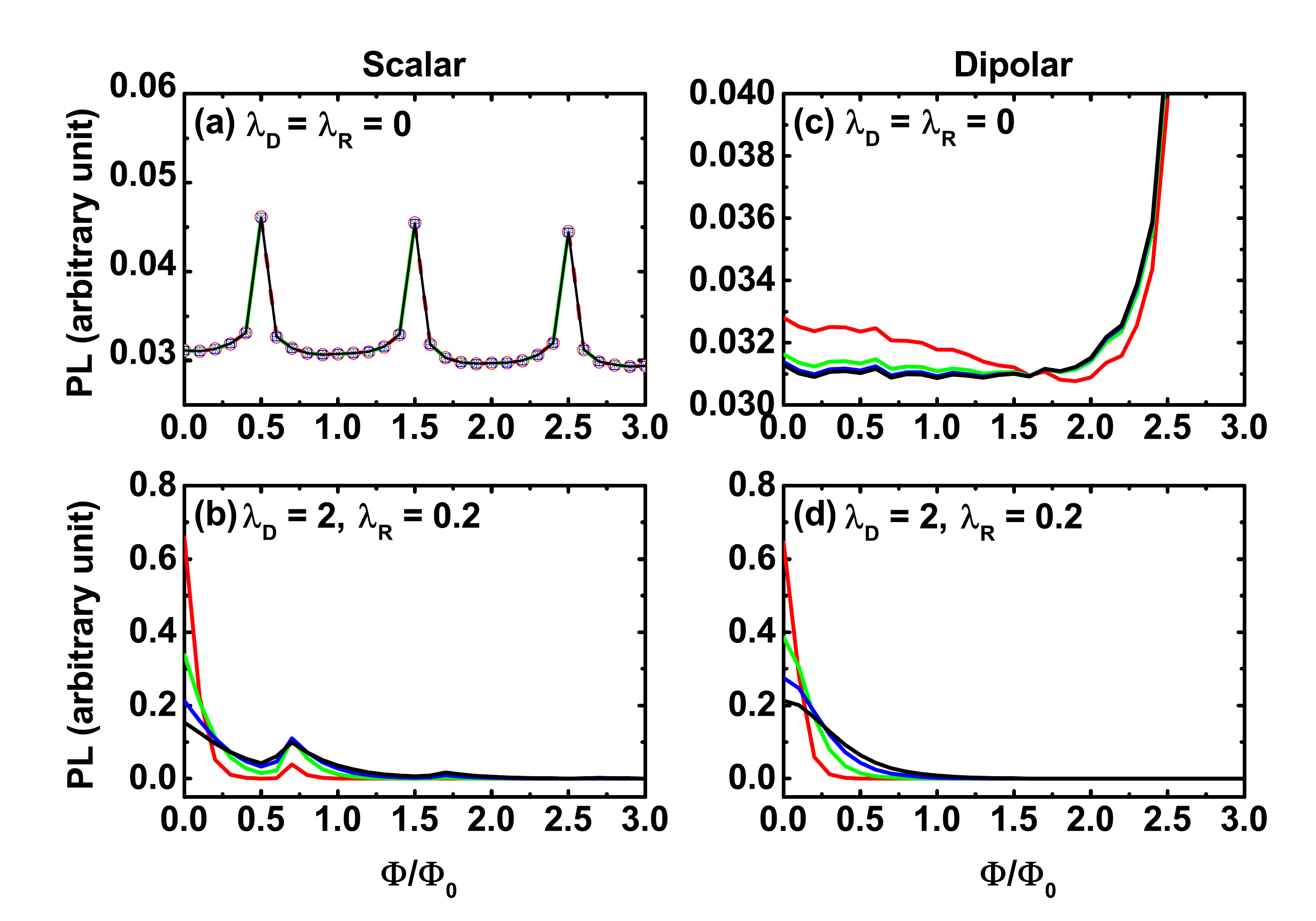}
\caption{(Color online)
The normalized magneto-PL spectra at finite temperatures, $0.05$\,K(red), $0.1$\,K(green), $0.15$\,K(blue), and $0.2$\,K(black). The left(right) panels display the
spectra of a scalar(dipolar) exciton, and for both cases the SOIs are excluded(included) in the upper(lower) figures. The strength parameters are set to $\lambda_D = 2$ and $\lambda_R = 0.2$,
in the unit of $\lambda^*$.
}
\label{Fig3}
\end{figure}

%
Fig.~\ref{Fig3} shows the normalized magneto-PL spectra at finite temperatures,
in which the left(right) panels display the spectra of an incoherently(coherently) moving exciton, and for both cases the upper(lower) ones include(exclude) the SOIs.
On the relevant energy scale of our calculations, a tiny thermal fluctuation can be enough to excite charge transitions via various open channels; thus it is indispensable to completely count all possible channels to achieve the converged evaluation of Eq.~(\ref{eqn:IT}).
But since highly excited eigenstates play insignificant roles in the determination of the AB property, the PL intensity mainly records photon emission of low energy exciton recombination.

In Fig.\ \ref{Fig3}(a), the variation of the PL intensity shows a periodic feature
with the same period as that of the lowest energy state of Fig.~\ref{Fig1}. Without the SOIs, the peaks right centered at $(n+1/2)\Phi/\Phi_0$ correspond to the occurrence of level crossings of states
with $L = 0$ and $J = 0$. Furthermore, there is a pseudo-AB feature valid for the ground state and a thermally excited bundle of states
with an approximately common Boltzmann weight leading to an undistinguishing thermal splitting of the PL spectra shown in (a).
On the other hand, a prominent thermal splitting and depression of the PL in (c) reflects the monotonic growth of the $L = 0$ bright exciton state
shown as black curves in Fig.~\ref{Fig2}(b). Accordingly, low-level avoided crossings explain
the rapid increase in the PL curves in the approach of $\Phi/\Phi_0 = 2.5$.
When the SOIs are involved, (b) and (d) reveal that the PL intensities
decrease rapidly with an increasing magnetic flux. This demonstrates a common tendency of a spin flip induced by the SOIs,
which turns the ground-state exciton into a diamagnet.
While the high PL intensities at zero magnetic field shown in (b) and (d) reflect the time-reversal and Kramer's 4-fold degeneracy enhancement,
a stronger enhancement in the absence of the SOIs at temperatures of 5 K and 10 K is observed.
{At these high temperatures, we further observe that the SOIs counteract the zero-field advantage of thermal repopulations, hence consequently resulting in the equal-field photon emission.}
Overall, we find that under the spatially uniform magnetic flux and light field, a coherent bright exciton can be sustained in the weak magnetic field regime. The presence of the SOIs is found to cause a blue shift of the bright region, but inevitably to depress the PL intensities at finite temperatures, and to stretch the dark regime.

%
%
%
%

\subsection{Persistent charge and spin currents}

\begin{figure}[thbp!]
\includegraphics[width=0.45\textwidth]{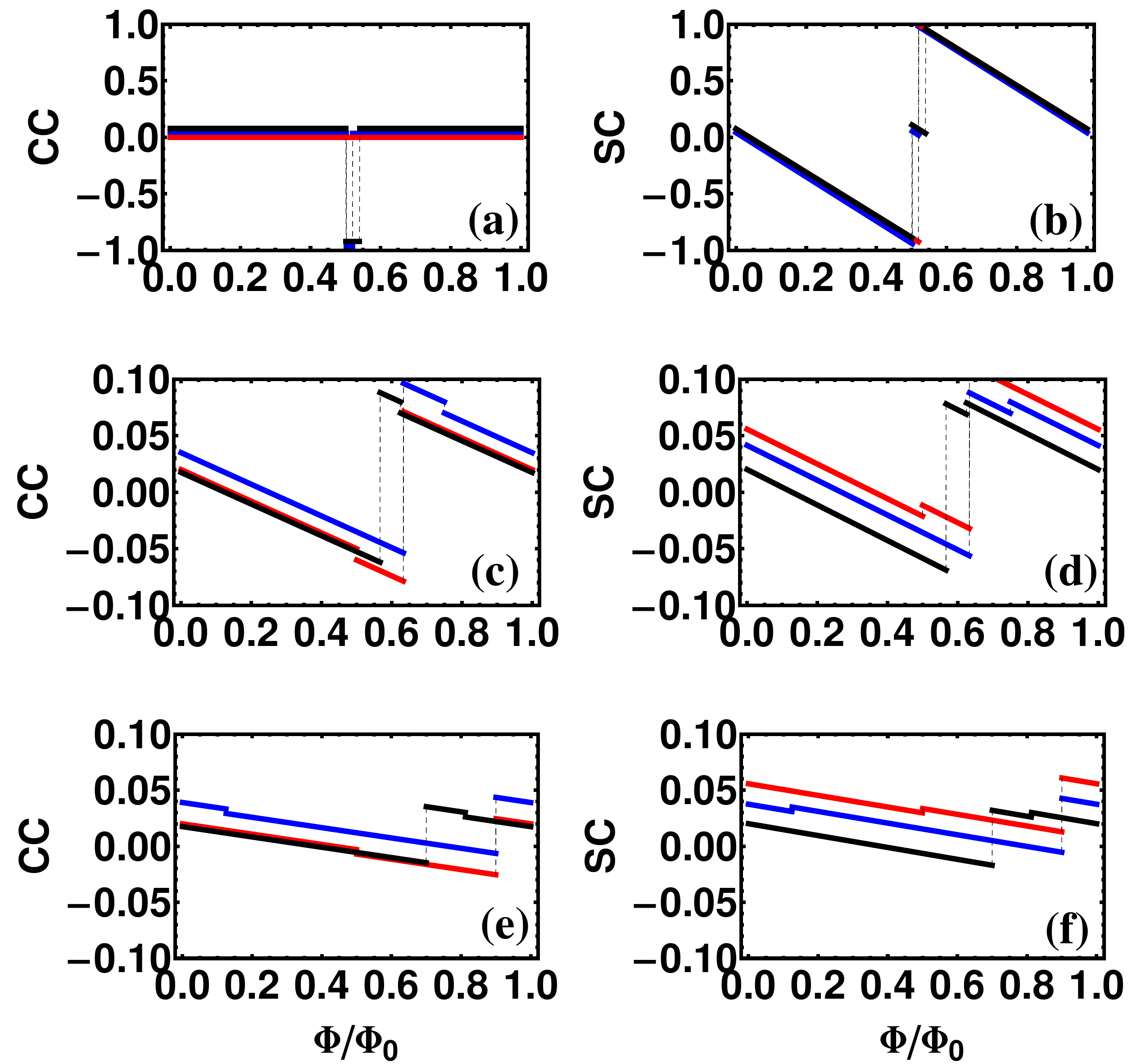}
\caption{(Color online)
Generation of the persistent CC and SC of a scalar exciton in different dimensionless sizes of rings and with different dimensionless SOI strengths in color lines of blue, red, and black. (a) and (b) describe a homomass exciton corresponding to $m^*_e = m^*_h$; $R_h \sim R_e = 1$; $\lambda_D = 0.14,\,0.14,\,0.2$; $\lambda_R = 0.05,\,0.14,\,0.035$. In (c) and (d), $m^*_e = 0.1\,m^*_h$; $R_h = 3.16$, $R_h = 2.58$; $\lambda_D = 0.14,\,0.14,\,0.1$; $\lambda_R = 0.05,\,0.14,\,0.035$. In (e) and (f),
$m^*_e = 0.1\,m^*_h$; $R_h = 3.16$, $R_h = 2.58$; $\lambda_D = 0.14,\,0.14,\,0.1$; $\lambda_R = 0.05,\,0.14,\,0.0035$.
}
\label{Fig4}
\end{figure}

As the PL intensity shows schematics of material structure and quality, nonzero equilibrium currents provide an alternative to probe coherent property of the ground state. Fig.~\ref{Fig4} demonstrates persistent CC and SC of a scalar exciton. Since the SOIs were found to reduce the Ising-type magnetic order by planar energy depletion, either the use of a tiny ring or weak SOIs may help to stabilize the system sustained solely by a rotational kinetic term in a hard-ring model.
The observation of single particle's AB oscillation is strong evidence that the wave functions of the charge carriers stay coherent along the ring circumference.
But once one charge moves diametrically opposed to the other, the two merge into a chargeless complex and eliminate the exciton's AB correlation.

First, we set a reference point by choosing a tiny ring with dimensionless radius $R_h \sim R_e = 1$ and letting $m^*_e = m^*_h$.
The CC and the SC with the dimensionless SOIs strengths given by $\lambda_D = 0.14,\,0.14,\,0.2$, $\lambda_R = 0.05,\,0.14,\,0.035$ corresponding to blue, red, and black curves are shown in {(a) and (b)}, in which the first pair are the corresponding parameters used in the
analysis of magneto-optical spectra.
As expected, when DSOI = RSOI, the electron and the hole hold equal but opposite angular momentum and cause a zero net CC. However, a different pattern of
the SC showing an instant flip occurs {indicating that the two spins becoming parallel}.
On the other hand, as the DSOI $\neq$ the RSOI, the appearance of plateaus is an indication of an asynchronous orbit shift of the individual charges.
{Occurring at spectra crossings, the alternating orbit shift turns the plateaus into intermediate states for the excitonic transitions
to be accomplished via two-photon processes}.

In order to mimic a realistic heavy-hole exciton time evolution, the free motion is restricted
by considering different finite sizes of confinement for the particles, by setting the dimensionless ring radius
$R_{h(e)} \sim 3.16(2.58)$ and $R_{h(e)} \sim 5(4.4)$. The modification leads to a generation of asymmetric and attenuated
CC and SC as is shown in {(c)-(d)} and {(e)-(f)}.
In these two cases, the third DSOI parameter is lowered to $0.1$ to preserve the validity of the perturbation condition.
{As in the band theory, the effective mass is introduced reflecting the curvature of the energy spectrum, the mass effects of two charged
particles also reflect their nonvanishing CC and SC}.
In the larger rings the zig-zag CC and SC are flattened. This is a direct revelation of the weakening of the circumferential rotations and the
diminishing of the single-particle AB features.

The sharp variation which occurs in the CC and the SC is an indication of the first order phase transition of the exciton states. In the first period, where $0 \leq \Phi/\Phi_0 \leq 1$, the phase transition is recognized as an exciton completing two-photon transitions, initially starting from $|J^e_z,s^e_z;J^h_z,s^h_z\rangle = |-1/2,-1/2;-1/2,-1/2\rangle$, passing over $|-3/2,-1/2;-1/2,-1/2\rangle$, and finally reaching $|-3/2,-1/2;1/2,-1/2\rangle$.
For the CC, a $\lambda$-transition picture is suitable to describe the imbalance. The orbit flip is completed by an absorption of a left circularly polarized gauge photon of the electron followed by an emission of a left circularly polarized gauge photon of the hole. Differing from the homomass exciton for which a stretch of bright PL regime can be established using the SOI discrepancy, it is found that the blue shift caused by the dominant DSOI tends to eliminate the bright PL of a heteromass exciton.
For the SC, a ladder-transition picture can be constructed via left and right circularly polarized gauge photon absorptions of the electron and the hole,
respectively. The two different patterns of the CC and the SC depict the complex system containing opposite charges but common spin orientation constituents.
While the RSOI is a weak perturbation, the hole ground state approaches close to a spin degeneracy. The zig-zag patterns unveil subsequent bright-dark-bright
switching via a gauge photon absorption and emission, which inspires the possibility to engineer an optical switch in a nano device by
adjusting the carrier concentrations.
%
%
%
%
%

\subsection{Persistent dipole and spin dipole currents}
\begin{figure}[thbp!]
\includegraphics[width=0.48\textwidth]{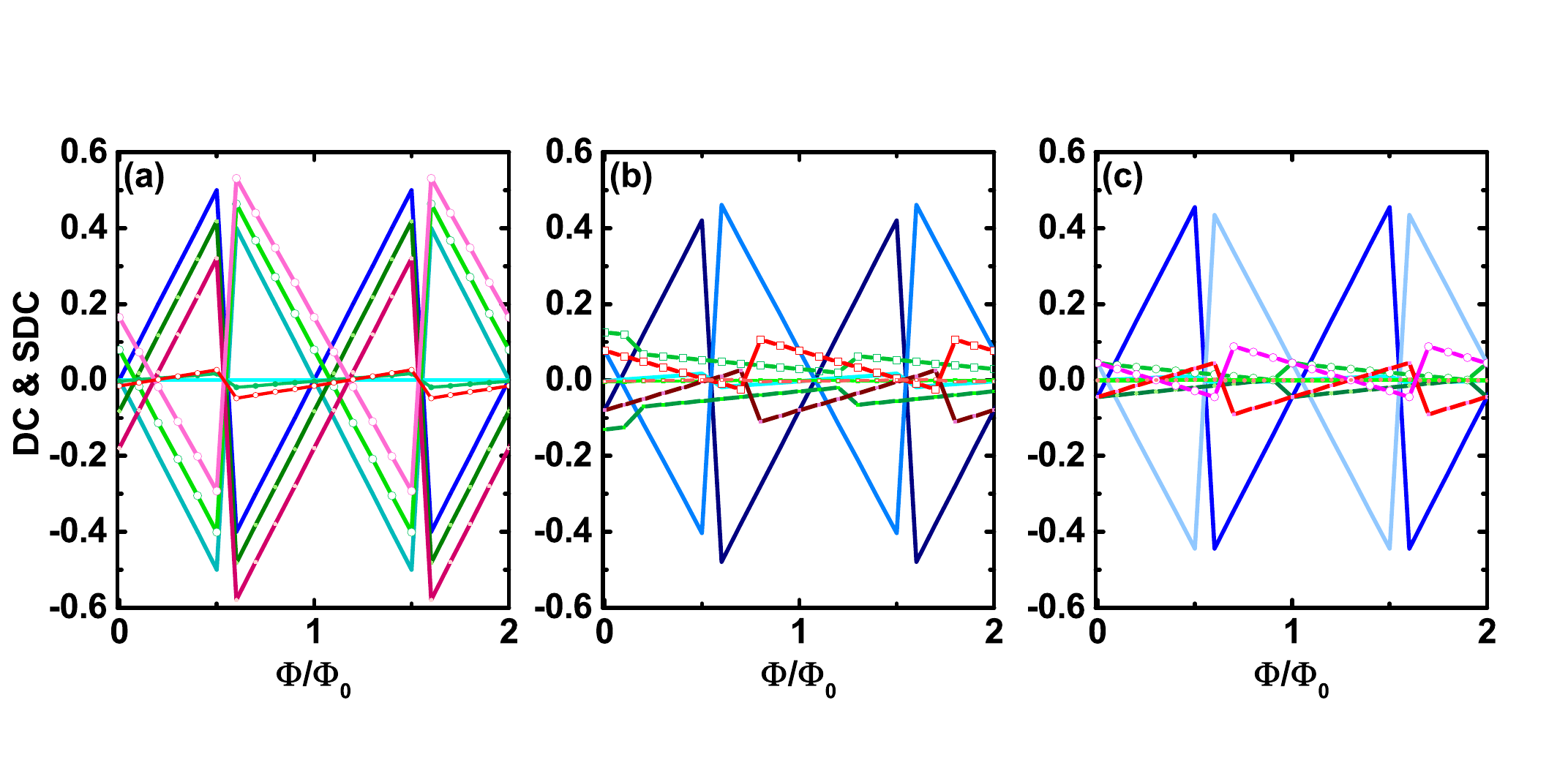}
\caption{(Color online)
Generation of DC and SDC of a dipolar exciton in different dimensionless sizes of rings and with different dimensionless SOI strengths in color lines ordering in blue, green, red for representing DC(dark), SDC(light), (DC+SDC)dash. (a) $R_d = 1$, $\lambda_+ = 0, 0.23, 0.3$, $\lambda_- = 0, 0.17, 0.26$; (b) $R_d = 1, 2.58, 4.4$, $\lambda_+ = 0.2, \lambda_- = 0$; and
(c) $R_d = 1, 2.58, 4.4$, $\lambda_+ = 0.15, \lambda_- = 0.13$.
}
\label{Fig5}
\end{figure}

Fig.~\ref{Fig5} demonstrates the DC(dark color lines) and the SDC(light color lines) in terms of dimensionless parameters
(a) $R_d = 1$, $\lambda_+ = 0, 0.23, 0.3$, and $\lambda_- = 0, 0.17, 0.26$, corresponding to $\lambda_D = 0, 0.2, 0.28$ and $\lambda_R = 0, 0.11, 0.1$;
%
(b) $R_d = 1, 2.58, 4.4$, $\lambda_+ = 0.2$, and $\lambda_- = 0$, corresponding to $\lambda_D = 0.14, \lambda_R = 0.14$;
and
(c) $R_d = 1, 2.58, 4.4$, $\lambda_+ = 0.15$, and $\lambda_- = 0.13$, corresponding to $\lambda_D = 0.14, \lambda_R = 0.05$, respectively,
which allow us to discuss the ring size and the SOI effects on the generation of the DC and the SDC.
The sawtooth oscillations of period $\Phi/\Phi_0$ characterize a single particle
feature for the dipolar exciton. The AB periodicity is analogously manifested by the alternating currents.
In larger rings and strong SOIs, the shift of the energy dispersion also causes a delay of rectification as a function of $\Phi/2\Phi_0$.
Due to the projected fractions $p_{\pm}$, the SDC are smaller than the DC, and the larger SOIs induce larger differences in the currents.
On the other hand, a general suppression of the currents was found in large-ring cases. This is because, as the ring is very large,
the rotational kinetic energy is too small in the low fields to support the zero angular momentum state. In some cases it would lead to the occurrence
of an anomalous AB effect demonstrated by a surge in currents as is shown in {the open-square green lines} of (b).
Analysis of the effects of the ring size and the SOIs on the generation of the DC and the SDC shows that the formation of a coherent dipolar exciton is
equivalent to a formation of a vortex, bringing in nonzero and robust angular momentum states to stabilize the system.

%
%
%
\subsection{Spin Orientation}

\begin{figure}[thbp!]
\includegraphics[width=0.8\textwidth]{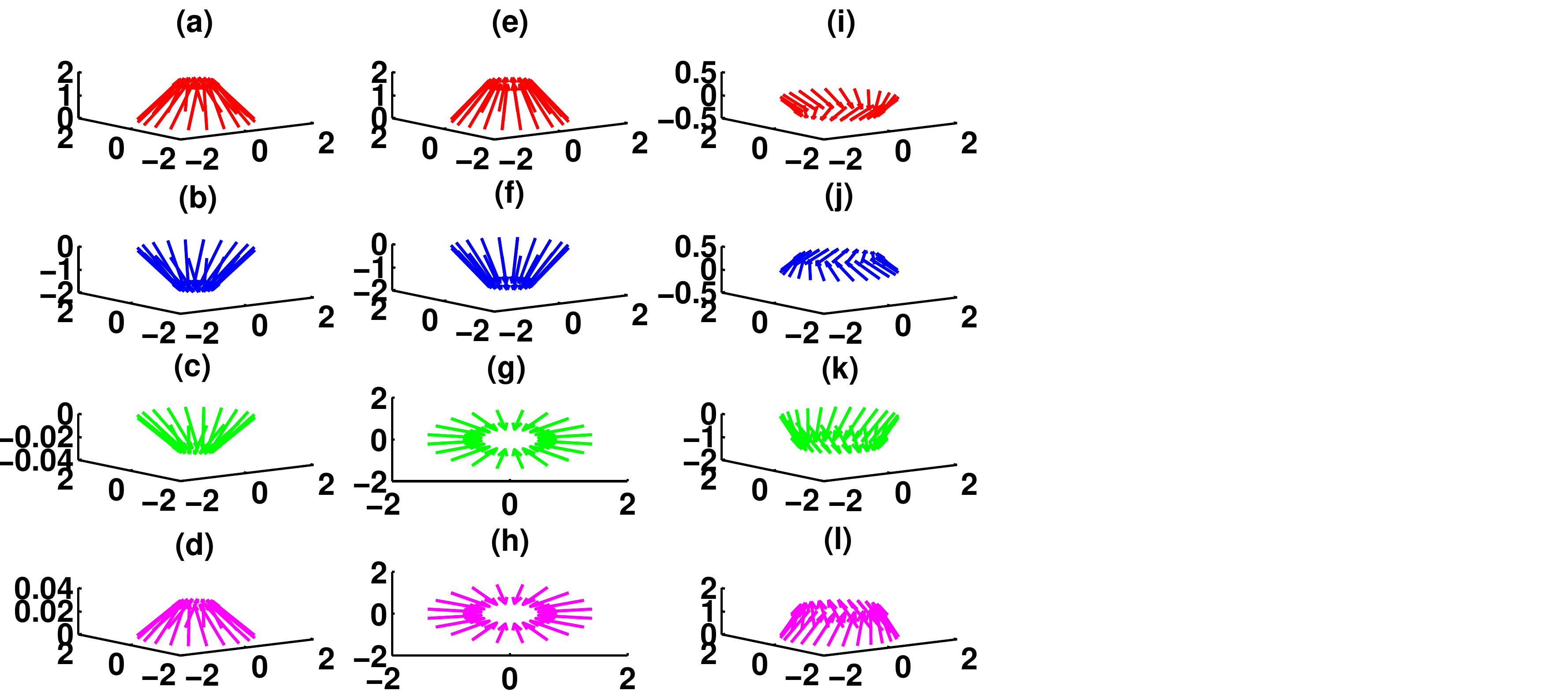}
\caption{(Color online)
Spin orientations of the scalar exciton with $\phi_e = -\phi_h$ and different dimensionless SOIs in terms of $\theta = \tan^{-1}\lambda_D$ and $\eta = \tan^{-1}\lambda_R$. In (a)-(d), $\theta = 0.14$, $\eta = 0.05$; in (e)-(h) $\theta = 0.14$, $\eta = 0.14$; and in (i)-(l) $\theta = 1.4$, $\eta = 0.5$.
}
\label{Fig6}
\end{figure}

\begin{figure}[thbp!]
\includegraphics[width=0.82\textwidth]{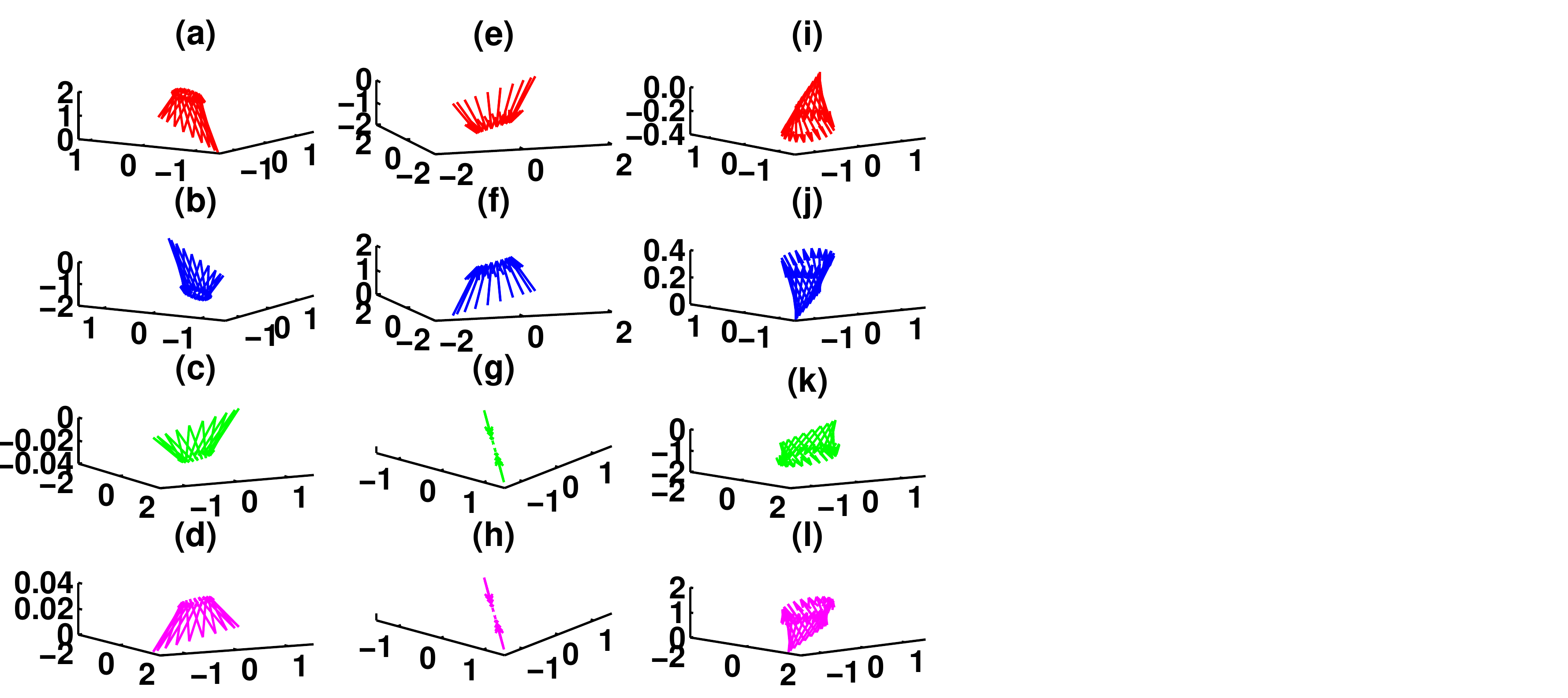}
\caption{(Color online)
Spin orientations of the dipolar exciton with $\phi_e = \phi_h$ and different dimensionless SOIs in terms of $\theta = \tan^{-1}\lambda_D$ and $\eta = \tan^{-1}\lambda_R$. In (a)-(d), $\theta = 0.14$, $\eta = 0.05$; in (e)-(h) $\theta = 1.4$, $\eta= 1.4$; and in (i)-(l) $\theta = 1.4$ and $\eta = 0.5$.
}
\label{Fig7}
\end{figure}

Spin textures provide an alternative probe to the influences of the SOIs. Based on  Eqs.~(\ref{Suu}-\ref{Sud}) and viewing from the electron's reference
of spatial coordinates, {Fig.~\ref{Fig6}} shows spin orientations of the {scalar exciton} when letting
$\phi_e = -\phi_h$, whereas {Fig.~\ref{Fig7}} displays the spin textures of the {dipolar exciton} with $\phi_e = \phi_h$.
In both figures, the spin orientations of $\uparrow\,\uparrow$, $\downarrow\,\downarrow$, $\uparrow\,\downarrow$, and $\downarrow\,\uparrow$ pairs are depicted in four rows, from up to down. The strength of the SOIs, from weak to strong, are listed in different columns and are represented in terms of the dimensionless SOIs by $\theta = \tan^{-1}\,\lambda_D = 0.14, 0.14, 1.4$\,rad, $\eta = \tan^{-1}\,\lambda_R = 0.05, 0.14, 0.5$\,rad for Fig.~{\ref{Fig6}} and
$\theta = \tan^{-1}\,\lambda_D = 0.14, 1.4, 1.4$\,rad, $\eta = \tan^{-1}\,\lambda_R = 0.05, 1.4, 0.5$\,rad for Fig.~{\ref{Fig7}}, respectively. The near critical behaviors
are discussed with parameters $\theta = \eta = 1.4$\,rad.

In contrast to the Zeeman effect that prefers to polarize atoms along the magnetic field, the RSOI and the DSOI modify the kinematics in the transverse plane. Therefore, the presence of the SOIs may break the transverse axial-symmetry and cause spatially inhomogeneous states of polarization.
Eqs.~(\ref{Suu}) and (\ref{Sud}) indicate that spin textures
are aligned along a ring of radius $\sqrt{\sin^2 2\eta + \sin^2 2\theta}$. For scalar excitons, the circularly polarized spin orientations demonstrate {the possibility of the generation of parallel flow of CC and SC}.
However, for a dipolar exciton, the additional cross weighting factors $\pm{|\sin 2\eta \sin 2\theta \sin 2\phi|}$ cause a spin breathing. This implies that
the SOIs have a great potential to modulate and turn the states of spin polarization of a dipolar exciton into linearly polarized and squeezed orientations.

Based on a thin-ring assumption, when both the DSOI and the RSOI provide off-diagonal torques that drive anti-circulating flows, the torsion provided by the RSOI is homogeneous, whereas it is sinusoidally varying when caused by the DSOI.
Therefore, for both scalar and dipolar excitons, as the DSOI = the RSOI, the dark excitons have fully polarized spin states, whereas a spin demagnetization
occurs in the bright excitons.
In the absence of a RSOI, we find that an increase of the DSOI leads to a spatial shrinkage of the spin textures, and a decrease in the spin polarization
until the system undergoes a phase transition from the paramagnetic state to the diamagnetic state.
Below the critical point of the DSOI, we also find that when the RSOI is present, for example when $\theta = 1.4$ and $\eta = 0.5$, a spin flip occurs in the dark excitons and a spin chain skewing is initiated for both the dark and the bright excitons. These can be understood since near the modulation boundary of $H_D$ the system is critical to the perturbation caused by $H_R$. While dominated by the $H_R$, the system acquires an energy to cross over the gap between the spin states and earns a torque to twist. By increasing $\lambda_R$ to a value larger than $\lambda_D$, a spin flip of the bright excitons is initiated.
Since Eqs.~(\ref{Suu}-\ref{Sud})
show periodic modulations in the spin orientations, by tuning the strength of the SOIs, we observe an evolution of spin density waves of these spin textures and the tunneling of Bloch walls formed by the SOIs.

\subsection{Optical Switches}

\begin{figure}[thbp!]
\includegraphics[width=0.40\textwidth]{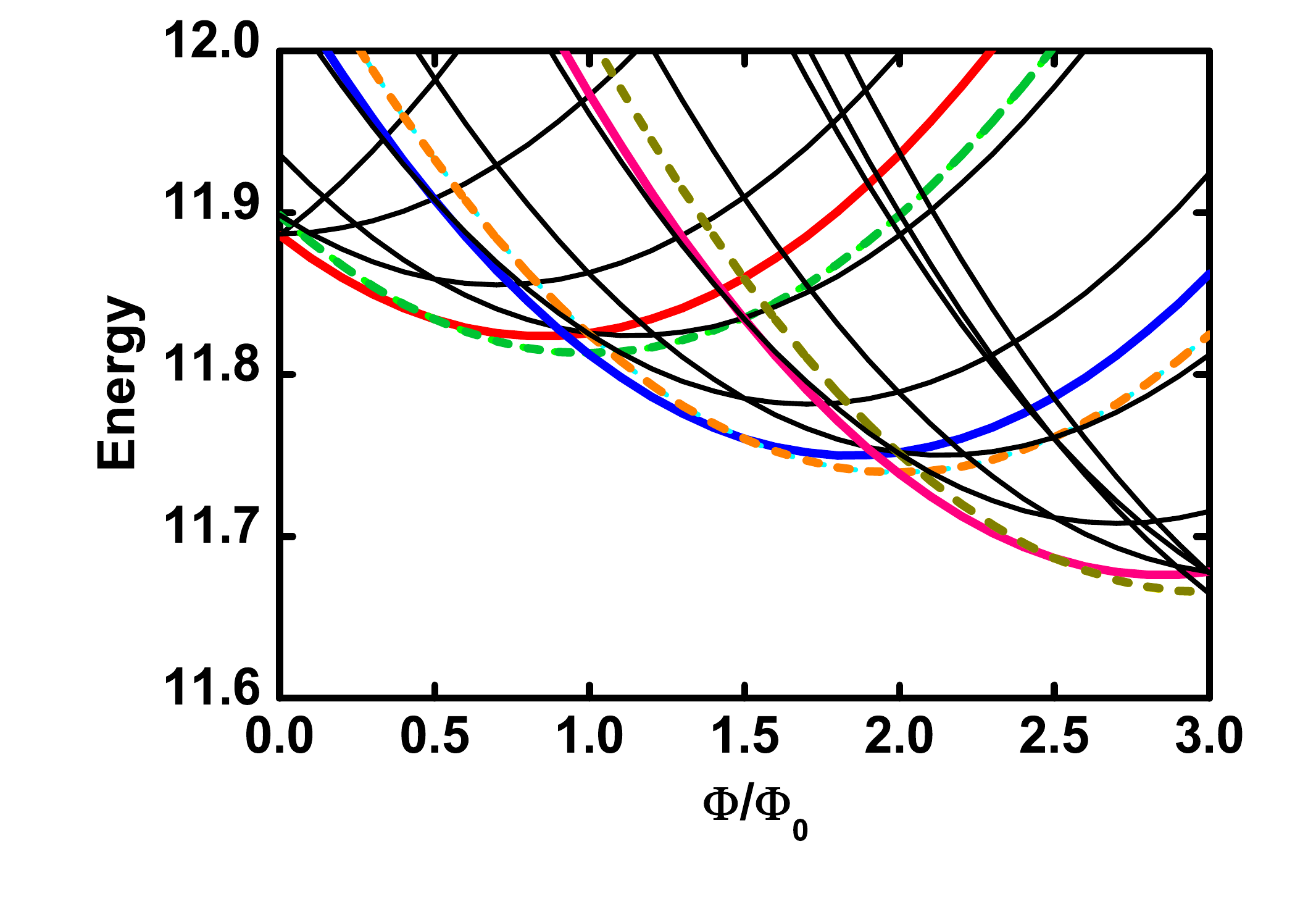}
\caption{(Color online)
The quantum switch of the spinor exciton. With $\lambda_D = 4\times 10^{-11}$\,eV-m and $\lambda_R = 0.05\times 10^{-11}$\,eV-m, the system has light-dark pair ground states colorfully specified for various angular momenta $|e\,h\rangle = |0\,0\rangle, |0\,1\rangle, |-1\,1\rangle, |-1\,2\rangle, |-2\,2\rangle$, and $|-2\,3\rangle$.
}
\label{Fig8}
\end{figure}

The success in the generation of pure spin currents in the homomass and scalar excitons with
$\lambda_D = \lambda_R$ inspires further investigation of the effects of the SOIs in the generation of pure CC and optical switching in our system.
According to {Eqs.~(\ref{Ie}) and (\ref{Ih}}), a thought experiment can be realized via adjusting the SOIs to critical values with
$\Theta_e = \Theta_h = \pi/2$ or via designing a highly unequal system with strong DSOI and weak RSOI to nearly approach a spin degeneracy.
In both approaches, nonzero electronic CC and excitonic CC can be established in scalar and dipolar excitons.
However, with large SOIs, the existence of a giant net spin dipole moments prohibits the recombination of the electron and the hole and a PL emission.
On the other hand, within a slanted potential, the excitonic ground states can be established with spin degeneracy to demonstrate
asynchronous blue shifts in the regime far below the critical Dresselhaus SOI.
As a result, a system with light-dark pair ground states, such as $|e\,h\rangle = |0\,0\rangle, |0\,1\rangle, |-1\,1\rangle, |-1\,2\rangle, |-2\,2\rangle$, and $|-2\,3\rangle$, (Fig.~\ref{Fig8}) can be adjusted to mimic an alternating optical device. The manipulation of the SOIs provides
a potential application for the quantum-rings spinor excitons to be utilized in nano-scaled magneto-optical switches.

%
%
%
%
%
%
\section{Conclusion}

In this work we have investigated the effects of the presence of the Rashba and the Dresselhaus spin-orbit interactions in the magneto-optical properties of
excitons in a two-dimensional semiconductor quantum double-ring threaded by a magnetic flux perpendicular to the plane of the rings.

 A scalar exciton influenced only by a magnetic flux through the
 holes of the rings in the absence of an effective magnetic field in the rings is found to move
 incoherently, while a dipolar exciton experiencing an effective magnetic
 field in the ring subjected to a flux through it will move coherently. Therefore,
 by adjusting the extent of the magnetic flux, an Aharonov-Bohm phenomenon
 can be observed for a dipolar exciton, producing an optical
 manifestation of a quantum interference.

In the presence of the SOIs, the light-matter interactions of the spinor exciton show that for scalar excitons, there are open channels for the
spontaneous recombination and leading to the bright
photoluminescence, whereas the forbidden recombination of dipolar excitons leads to the dark photoluminescence.

We have investigated the generation of persistent charge and spin currents.
From the ladder- and $\lambda$-like piecewise continuous and linear functions of CC and SC we can distinguish the electrical
and spin properties of charge carriers. We can determine the routes of the excitonic transitions that are signaled by
the absorption and the emission of polarized photons.

By calculating the two-particle spin orientations under various spin-orbit interactions, we found that
the SOIs induce off-diagonal torques and drive anti-circulating flows. As the RSOI has the potential to initiate a homogeneous
spin chain skewing, the DSOI provides a sinusoidally modulated torque leading a phase transition from the paramagnetic state to a diamagnetic state.
As a consequence, the scalar exciton demonstrates a circularly polarized spin distribution whereas the dipolar exciton demonstrates
itself as a squeezed complex with a specific spin polarization.
Moreover, a coherently moving dipolar exciton acquires a nontrivial dual Aharonov-Casher phase, bringing the possibility to generate persistent
dipole currents, and the spin dipole currents.
By tuning the strength of the SOIs, we observe an evolution of the spin density waves of these spin textures and
the tunneling of Bloch walls formed by the SOIs. The analysis
of the generation of the DC and the SDC shows that the formation of a coherent dipolar exciton is
equivalent to the formation of a vortex, bringing in nonzero and robust angular momentum states to stabilize the system.

Finally, we have shown that within a slanted potential, excitonic ground states can be established with spin degeneracy in the regime
far below the critical DSOI. This implies that a system with sequential light-dark pair ground states can be adjusted to mimic an alternating
optical device. Our study reveals that the manipulation of the spin-orbit interactions provides a potential application for the
quantum-ring spinor exciton to be utilized in nano-scaled magneto-optical switches.

\section*{Acknowledgements}
We would like to thank the financial support by the Ministry of Science and Technology, Taiwan, with grant MOST 104-2511-S-845-009-MY3,
the Icelandic Research Fund, grant no.\ 163082-051, and the Research Fund of the University of Iceland.

%
%
%
%
\bibliographystyle{apsrev4-1}
%

%
%
%
%

\end{document}